\documentclass[acmsmall,screen]{acmart}

\acmJournal{JACM}
\acmVolume{37}
\acmNumber{4}
\acmArticle{111}
\acmYear{2018}
\acmMonth{8}
\setcopyright{none}
\renewcommand\footnotetextcopyrightpermission[1]{}

\usepackage{tikz}
\usetikzlibrary{arrows.meta,calc,fit,backgrounds,positioning}
\usepackage{fontawesome5}

\usepackage{enumitem}
\usepackage{url}
\usepackage{multirow}
\usepackage{makecell}
\usepackage{amsmath}
\usepackage{algorithm}
\usepackage{algpseudocode}
\newcommand{\todo}[2][]{}
\newcommand{\missingfigure}[2][]{}

\usepackage{colortbl}
\usepackage{xspace}
\usepackage{color}
\usepackage{listings}
\usepackage[many]{tcolorbox}
\usepackage{lipsum}

\usepackage{pifont}
\usepackage{booktabs}
\usepackage{fancyvrb}

\definecolor{pygFun}{rgb}{0.05,0.20,0.55}   
\definecolor{pygAtom}{rgb}{0.05,0.45,0.10}  
\definecolor{pygKw}{rgb}{0.60,0.10,0.10}    
\definecolor{pygPunc}{rgb}{0.40,0.40,0.40}  
\definecolor{pygCom}{rgb}{0.50,0.50,0.50}   
\definecolor{cOrange}{HTML}{E69F00}
\definecolor{softyellow}{RGB}{255,250,205}   
\definecolor{paleyellow}{RGB}{255,255,224}    
\definecolor{beigeellow}{RGB}{255,248,220}    
\definecolor{creamyellow}{RGB}{255,253,208}  
\makeatletter
\@namedef{pyg@@n+nf}#1{\textcolor{pygFun}{#1}}
\@namedef{pyg@@l+s+sAtom}#1{\textcolor{pygAtom}{#1}}
\@namedef{pyg@@k}#1{\textcolor{pygKw}{#1}}
\@namedef{pyg@@p}#1{\textcolor{pygPunc}{#1}}
\@namedef{pyg@@c}#1{\textcolor{pygCom}{\textit{#1}}}
\newcommand{\PYG}[2]{\@ifundefined{pyg@@#1}{#2}{\@nameuse{pyg@@#1}{#2}}}
\makeatother
\newcommand{\PYGZus}{\textunderscore}

\DefineVerbatimEnvironment{MintedVerbatim}{Verbatim}{fontsize=\footnotesize}

\newtcolorbox{findingbox}{
    colback=gray!10,
    colframe=gray!50,
    boxrule=0.5pt,
    arc=1mm,
    boxsep=1.5mm,
    left=1.7mm,right=1.7mm,top=1mm,bottom=1mm,
    fontupper=\itshape
}

\definecolor{noticegold}{RGB}{205,170,40}
\newtcolorbox{noticebox}{
    colback=white,
    colframe=cOrange,
    boxrule=0.5pt,
    arc=1mm,
    boxsep=1.5mm,
    left=1.7mm,right=1.7mm,top=1mm,bottom=1mm
}

\newtcolorbox{notebox}[1][]{
colback=pygAtom!15, 
colframe=pygAtom!60,
    boxrule=0.5pt,
    arc=1.5mm,
    boxsep=1.2mm,
    left=2mm,right=2mm,top=2mm,bottom=2mm,
    #1
}
\newtcolorbox{answerbox}{
  enhanced,
  left=1.7mm,
  right=1.7mm,
  top=1.7mm,
  bottom=1.7mm,
  colback=gray!10,  
  colframe=gray!90, 
  boxrule=0pt,      
  leftrule=3pt,     
  sharp corners,    
  breakable         
}

\newcounter{findingcounter}
\newcommand{\finding}[1]{%
    \stepcounter{findingcounter}%
    \begin{tcolorbox}[colframe=black, boxrule=0.8pt, arc=1mm]
    \textbf{Finding \thefindingcounter:} #1
    \end{tcolorbox}
}

\definecolor{verylightgray}{rgb}{.97,.97,.97}
\lstdefinelanguage{Solidity}{
  keywords=[1]{anonymous, assembly, assert, balance, break, call, callcode, case, catch, class, constant, continue, constructor, contract, debugger, default, delegatecall, delete, do, else, emit, event, experimental, export, external, false, finally, for, function, gas, if, implements, import, in, indexed, instanceof, interface, internal, is, length, library, log0, log1, log2, log3, log4, memory, modifier, new, payable, pragma, private, protected, public, pure, push, require, return, returns, revert, selfdestruct, send, solidity, storage, struct, suicide, super, switch, then, this, throw, transfer, true, try, typeof, using, view, while, with, addmod, ecrecover, keccak256, mulmod, ripemd160, sha256, sha3}, 
  keywordstyle=[1]\color{black}\bfseries,
  keywords=[2]{address, bool, byte, bytes, bytes1, bytes2, bytes3, bytes4, bytes5, bytes6, bytes7, bytes8, bytes9, bytes10, bytes11, bytes12, bytes13, bytes14, bytes15, bytes16, bytes17, bytes18, bytes19, bytes20, bytes21, bytes22, bytes23, bytes24, bytes25, bytes26, bytes27, bytes28, bytes29, bytes30, bytes31, bytes32, enum, int, int8, int16, int24, int32, int40, int48, int56, int64, int72, int80, int88, int96, int104, int112, int120, int128, int136, int144, int152, int160, int168, int176, int184, int192, int200, int208, int216, int224, int232, int240, int248, int256, mapping, string, uint, uint8, uint16, uint24, uint32, uint40, uint48, uint56, uint64, uint72, uint80, uint88, uint96, uint104, uint112, uint120, uint128, uint136, uint144, uint152, uint160, uint168, uint176, uint184, uint192, uint200, uint208, uint216, uint224, uint232, uint240, uint248, uint256, var, void, ether, finney, szabo, wei, days, hours, minutes, seconds, weeks, years},  
  keywordstyle=[2]\color[rgb]{0.3,0.5,0.3}\bfseries,
  keywords=[3]{block, blockhash, coinbase, difficulty, gaslimit, number, timestamp, msg, data, gas, sender, sig, now, tx, gasprice, origin},  
  keywordstyle=[3]\color{violet}\bfseries,
  identifierstyle=\color{black},
  sensitive=false,
  comment=[l]{//},
  morecomment=[s]{/*}{*/},
  commentstyle=\color{gray}\ttfamily,
  stringstyle=\color{red}\ttfamily,
  morestring=[b]',
  morestring=[b]"
}

\definecolor{backgray}{gray}{0.95}

\definecolor{prompt}{RGB}{25,78,132}

\tcbset{
  normal/.style={
    colframe=black,
    coltitle=white,
    colback=white,
    colbacktitle=black!70,
    fonttitle=\bfseries,
    boxrule=0.8pt,
    arc=3pt,
    left=6pt,
    lefttitle=6pt,
    right=6pt,
    righttitle=6pt,
    top=6pt,
    bottom=6pt,
    toptitle=3pt,
    bottomtitle=3pt,
    title style={left color=black!70, right color=black!70},
  }
}

\tcbset{
  prompt/.style={
    colframe=black,
    coltitle=white,
    colback=white,
    colbacktitle=prompt,
    fonttitle=\bfseries,
    boxrule=0.8pt,
    arc=3pt,
    left=6pt,
    lefttitle=6pt,
    right=6pt,
    righttitle=6pt,
    top=6pt,
    bottom=6pt,
    toptitle=3pt,
    bottomtitle=3pt,
    title style={left color=prompt, right color=prompt},
  }
}

\newcommand{\addr}[1]{{\small\path{#1}}}

\newcommand{\paratitle}[1]{\vspace{0.3em}\noindent\textbf{#1.}\quad}
\newcommand{\etal}{{\textit{et al.}}\xspace}
\newcommand{\eg}{{\textit{e.g.}}\xspace}
\newcommand{\ie}{{\textit{i.e.}}\xspace}

\newcounter{cnum}
\newsavebox{\griefbox}
\newcommand{\tool}{\textsc{GhostHunter}\xspace}

\newcommand{\ghostFillCount}{2{,}037{,}304\xspace}
\newcommand{\ghostFillParticipants}{261{,}156\xspace}
\newcommand{\collateralAtRisk}{\$1.83\,B\xspace}
\newcommand{\peakDailyRate}{8.5\%\xspace}
\newcommand{\peakHourlyRate}{24.3\%\xspace}
\newcommand{\operatorGasBurned}{2.36\,M\xspace}

\newcommand{\avoidedLoss}{\$103.7\,M\xspace}
\newcommand{\uniqueOrders}{539{,}843\xspace}

\newcommand{\nonceBumpCount}{25{,}790\xspace}
\newcommand{\balanceDrainCount}{143{,}062\xspace}
\newcommand{\allowanceRevokeCount}{29{,}716\xspace}
\newcommand{\proxyTrapCount}{795{,}628\xspace}
\newcommand{\attackAttributedCount}{994{,}196\xspace}
\newcommand{\attackAttributedPct}{48.8\%\xspace}
\newcommand{\attackCollateral}{\$1.46\,B\xspace}
\newcommand{\attackGasBurned}{2.17\,M\xspace}
\newcommand{\proxyTrapGasBurned}{2.16\,M\xspace}
\newcommand{\pct}[1]{#1\%}

\newcommand{\reusedContracts}{167\xspace}
\newcommand{\reuseChainCount}{10\xspace}
\newcommand{\reuseActive}{31\xspace}
\newcommand{\reuseAvgTx}{369\,K\xspace}
\newcommand{\jaccardOneCount}{71\xspace}

\newcommand{\studyStart}{Aug 15, 2025\xspace}
\newcommand{\studyEnd}{Aug 15, 2026\xspace}
\newcommand{\vTwoCutover}{Apr 28, 2026\xspace}

\begin{document}

\title{The Ghosts of Polymarket: When Off-Chain Matches Meet On-Chain Reverts}

\authorsaddresses{}
\author{Yiming Shen}
\affiliation{\institution{Sun Yat-sen University}\country{China}}
\affiliation{\institution{Zhejiang University}\country{China}}
\author{Yuhan Jin}
\affiliation{\institution{Zhejiang University}\country{China}}
\author{Shuohan Wu}
\affiliation{\institution{Zhejiang University}\country{China}}
\author{Yanlin Wang}
\affiliation{\institution{Sun Yat-sen University}\country{China}}
\author{Jiachi Chen}
\affiliation{\institution{Zhejiang University}\country{China}}

\begin{abstract}
Polymarket has emerged as a prominent prediction market platform and one of the fastest-growing applications in DeFi.
To achieve low-latency trading, it adopts a hybrid architecture that matches orders off-chain but settles them on-chain for final execution.
This design creates a consistency gap we call \textit{Ghost Fills}: an order that is successfully matched off-chain may later fail during on-chain settlement. 
To understand the security implications of this gap, we investigate such failed settlements by building \tool, which reconstructs them from on-chain traces and attributes them to concrete attack patterns.
Across \ghostFillCount reverted match-order transactions, we find that attackers exploit the time gap between matching and settlement to invalidate already matched orders before they are finalized on-chain.
We identify four attack vectors, Nonce Bump, Balance Drain, Allowance Revoke, and Proxy Trap, realized via 37 evolving variants.
These vectors account for \attackAttributedCount of the reverts, enabling risk-free prediction, arbitrage-bot hunting, and liquidity reward manipulation. The cancelled orders spare their owners \avoidedLoss in net losses that settlement would have locked in, and the attacks cost the operator \attackGasBurned POL (about \$162\,K) in gas. During peak hours, up to \peakHourlyRate of transactions reverted, causing de facto DoS attacks. We also find that code derived from the flawed contract still appears in \reusedContracts independent contracts across \reuseChainCount chains holding at least \$23\,M in user funds, extending the impact beyond Polymarket. We have disclosed our evidence to affected parties, and the issue has been partially mitigated.

\end{abstract}

\ccsdesc[500]{Security and privacy~Software and application security}


\maketitle

\section{Introduction}
\label{sec:introduction}

Over the past few years, Polymarket~\cite{polymarketPolymarketWorldsLargest2026} has emerged as a breakout Web3 application, attracting broad public attention by allowing users to trade on the outcomes of real-world events. More broadly, it illustrates the potential of Web3 to move beyond token-centric speculation and support concrete, user-facing applications. By the time of writing, Polymarket had grown to more than \$450M in total value locked (TVL), over 100K daily active addresses~\cite{defillamaPolymarketTVLFees2026}, and about 1.75M on-chain settlement transactions per day on average over our study window, peaking at 4.5M (Figure~\ref{fig:rq1-reverts-daily}). To sustain this volume while keeping low-latency trading, Polymarket adopts a hybrid architecture that maintains an off-chain central limit order book (CLOB) for fast matching and uses on-chain settlement on Polygon for final execution~\cite{polymarketPolymarketDocumentation2026}. Through this split execution model, Polymarket preserves the trust guarantees of decentralized settlement while achieving the performance of conventional Web services.

However, this architecture with split execution leaves a critical window between off-chain matching and on-chain settlement: \textit{an order can appear filled off-chain while its corresponding settlement transaction has not, or may never be finalized at all.} Users, AI agents~\cite{xiaoTradingAgentsMultiAgentsLLM2025}, and trading bots~\cite{niedermayerDetectingFinancialBots2024} observing the fill through Polymarket's UI or CLOB API~\cite{polymarketPolymarketDocumentation2026} may treat it as a completed trade, even though the on-chain transaction remains pending and may still revert. Acting on such a fill leaves them exposed to adverse price movements in the interval. We refer to this phenomenon as \textit{Ghost Fills}.
This failure is especially damaging in prediction markets that host many short-cycle contracts (\eg, 5-min BTC price prediction~\cite{polymarket5MinuteCryptoOdds2026}), where the value of a fill can shift within seconds as external information arrives. A \textit{Ghost Fill} thus shifts the risk of delayed settlement onto the party that trusted the reported fill.

This emergence of \textit{Ghost Fills} raises questions: are they merely isolated order failures? Do they expose a structural vulnerability in the boundary between off-chain matching and on-chain settlement? Moreover, given the open-source nature of blockchain, new projects often inherit or fork existing code (\ie, \textit{contract reuse})~\cite{wangCopyandPasteIdentifyingEVMInequivalent2025}; does such reuse propagate the vulnerability to a broader ecosystem?
To understand the broader risks they reveal, we focus on three research questions:

\begin{itemize}[leftmargin=*]
    \item \textbf{RQ1.} How prevalent and costly are \textit{Ghost Fills}?
    \item \textbf{RQ2.} How do attackers exploit \textit{Ghost Fills}?
    \item \textbf{RQ3.} How far do \textit{Ghost Fill} risks spread through contract reuse?
\end{itemize}

To answer these questions, we build \tool, a trace-based measurement pipeline for \textit{Ghost Fills}. It first collects reverted \texttt{matchOrders} transactions from Polygon~\cite{polygonscan.comPolygonPoSChain2025} and maps each on-chain failure surface. We then classify \textit{Ghost Fills} on them to measure their prevalence, cost, and affected markets. To understand why these failures occur, \tool further performs causal failure analysis by examining the on-chain evidence associated with each failure surface. This analysis separates incidental failures from deliberate cancellations and uncovers four attack types with 37 variants that exploit this settlement window for profit. Finally, to assess whether the same risk extends beyond Polymarket, \tool scans verified contracts across 401 chains that reuse Polymarket-like exchange designs. We have the following major findings:

\noindent $\bullet$
\textbf{RQ1: \textit{Ghost Fills} are widespread and costly.}
From \studyStart{} to \studyEnd, \tool identifies \ghostFillCount reverted \texttt{matchOrders} transactions involving \ghostFillParticipants distinct participants. Their daily rate rose sharply in early 2026 and peaked at \peakDailyRate of all settlements. These failures are dominated by empty collateral balances and rejected token-delivery callbacks and affect fills involving \collateralAtRisk of collateral while burning \operatorGasBurned POL in operator gas. \textit{Ghost Fills} are a systemic and costly failure mode of Polymarket’s hybrid settlement design.

\noindent $\bullet$ 
\textbf{RQ2: Attackers exploit four evolving vectors.}
\tool attributes \attackAttributedCount of the \ghostFillCount \textit{Ghost Fills} (\attackAttributedPct) to attacks spanning Nonce Bump, Balance Drain, Allowance Revoke, and Proxy Trap. These attacks involve 37 implementation variants, reflecting an active cat-and-mouse cycle against Polymarket’s patches and monitoring. Together, they place \attackCollateral of collateral at risk and burn \attackGasBurned POL in operator gas~(\$162\,K), accounting for \pct{92} of all gas burned by reverts. At their May 4 peak, they drove the hourly revert rate to \peakHourlyRate of \texttt{matchOrders} transactions, creating a de facto DoS.

\noindent $\bullet$ 
\textbf{RQ3: \textit{Ghost Fill} risk has spread across chains.}
After scanning 32{,}103{,}371 Sourcify-verified contracts, \tool finds \reusedContracts independent Polymarket-like deployments across \reuseChainCount chains, including \jaccardOneCount copies with an identical function-selector set, holding at least \$23\,M in user funds. We further confirm \textit{Ghost Fills} on four of these deployments, including the two largest. Reuse alone does not make the gap exploitable: the same settlement gap remains largely harmless in NFT trading, but becomes profitable in prediction markets where a reported fill is worth voiding.

We further analyze how these attacks operate in practice. Attackers use the \textit{Ghost Fill} window for risk-free prediction, arbitrage-bot hunting, and liquidity-reward~\cite{polymarketLiquidityRewards2026} manipulation. Because these gains are hard to trace to addresses, we measure the loss each cancellation avoids: had the attacked orders settled, their owners would have lost \avoidedLoss net, 73 cents per dollar of notional. These activities are concentrated among a small number of adversaries acting at scale. They use Sybil strategies to mass-produce disposable wallets and launder the funds used to pay gas, weakening per-account blacklisting. We reported our evidence to Polymarket through three rounds of responsible disclosure. Polymarket has already shipped mitigations~\cite{polymarketMigratingCLOBV22026,polymarketDepositWallets2026}. However, these mitigations remain constrained by the timing gap in the hybrid architecture: an off-chain fill remains provisional until its on-chain settlement succeeds.
As of this writing, \textit{Ghost Fills} persist on Polymarket.

This paper makes three contributions:

\begin{itemize}
    \item We reveal \textit{Ghost Fills} as a security failure mode in hybrid prediction markets and show how attackers deliberately induce them through \textit{Cancellation Attacks}, across four attack vectors: Nonce Bump, Balance Drain, Allowance Revoke, and Proxy Trap.
    \item We build \tool, a trace-based analysis engine that measures the prevalence, cost, timing, affected markets, and attacker behavior behind \textit{Ghost Fills}. With \tool we attribute \attackAttributedCount reverts to \textit{Cancellation Attacks}, catalog 37 implementation variants across the four vectors, and quantify their impact on Polymarket users and settlement operators.
    \item We show that \textit{Ghost Fill} risk spreads through contract reuse, finding \reusedContracts reused contracts across \reuseChainCount chains. We responsibly disclosed our evidence to the affected third parties, and we release the artifacts needed to reproduce our measurements~\footnote{\url{https://github.com/shenyimings/ghost-hunter}}.
\end{itemize}

\section{Background}
\label{sec:background}

\subsection{Prediction Markets}
\label{sec:bg:prediction-markets}

A prediction market lets participants trade contracts whose payoff is tied to a future event, so that a contract's live price reads as the crowd's implied probability of that event~\cite{wikipediaPredictionMarket2026}. Polymarket pioneered this design at Web3 scale and remains the largest venue of its kind~\cite{defillamaPolymarketTVLFees2026}. Like any liquid exchange, it sustains an ecosystem of specialized actors around the order flow~\cite{huangOverviewWeb3Technology2024a}: retail traders take directional positions, market makers quote both sides, and arbitrage bots keep related contracts consistent. All of them act on the same public feed of reported fills. A fill that is reported but never settles therefore harms not only its counterparty but every actor that already traded on it.

\subsection{Polymarket Design}
\label{sec:bg:polymarket}

\paratitle{Hybrid architecture and order lifecycle} Polymarket splits a trade across two layers. Matching occurs off-chain in a central limit order book (CLOB)~\cite{polymarketPolymarketDocumentation2026}, while settlement occurs on-chain on Polygon~\cite{polygonscan.comPolygonPoSChain2025}. A user signs an order off-chain and submits it to the CLOB, which checks the signature, balance, and allowance before listing the order. When the crossing orders arrive, the CLOB matches them, reports the \emph{fill} to both sides immediately~\cite{polymarketPricesOrderbook2026}, and queues the match for settlement. A Polymarket-operated account, the \emph{operator}, then submits an on-chain \texttt{matchOrders} transaction to the Exchange contract, which verifies the signatures and transfers collateral and outcome tokens~\cite{polymarketCtfexchange2026}. Settlement is atomic: either the whole trade commits, or the transaction reverts and on-chain state is unchanged~\cite{polymarketPricesOrderbook2026}.

\paratitle{Outcome tokens} The asset transferred at settlement is an \emph{outcome token}, minted by the Conditional Tokens Framework (CTF)~\cite{polymarketCtfexchange2026} as an ERC1155 token on Polygon~\cite{ethereum.orgERC1155MultiTokenStandard2026}. A binary market has two (\textsc{Yes} and \textsc{No}), and the winning token redeems one-for-one against collateral once the event resolves through the UMA optimistic oracle~\cite{eskandariSoKOraclesGround2021}. Markets with more than two mutually exclusive outcomes (\ie, NegRisk) settle through a separate \emph{NegRisk CTF Exchange} contract~\cite{polymarketCtfexchange2026}, which our measurement also covers. ERC1155 also lets a recipient refuse a delivery. When tokens go to a contract, the token contract calls the recipient's \texttt{onERC1155Received} hook and completes the transfer only if the hook returns a fixed acceptance value. If the hook reverts or returns anything else, the transfer reverts~\cite{ethereum.orgERC1155MultiTokenStandard2026}, and so does any settlement that contains it.

\paratitle{Accounts and order signing} Polymarket is non-custodial: a user's funds stay in an account the user controls, whether a plain EOA or a smart-contract wallet such as a Polymarket proxy or a Gnosis Safe~\cite{gnosisMultisigWalletSecure}, and never in Polymarket's hands. The protocol separates two roles that the contract tracks independently. The \emph{owner} of an order is the wallet whose collateral and outcome tokens back it (the contract's \texttt{maker} field); the \emph{signer} is the key that signs it, which may be the owner itself or a delegate authorized to act for it. A match pairs one \emph{taker order}, which crosses the book, with one or more resting \emph{maker orders}. Following trading convention, we call their owners the \emph{taker} and the \emph{makers}. A Gnosis Safe is a thin proxy contract. Each wallet stores its own owners and settings but delegates all logic to a shared implementation contract called the \emph{singleton}. The singleton does not implement token-receiver hooks such as \texttt{onERC1155Received}. It forwards any call it does not recognize to a \emph{fallback handler}, whose address the wallet keeps in a fixed storage slot~\cite{gnosisMultisigWalletSecure}. The default handler accepts every delivery, but the owners can choose a different handler when they create the wallet or replace it later through \texttt{setFallbackHandler}.

\subsection{On-Chain Settlement on Polygon}
\label{sec:bg:settlement}

\paratitle{Transactions, contracts, and gas} Polymarket settles on Polygon PoS, an EVM blockchain that orders transactions into blocks roughly every two seconds~\cite{polygonscanPolygonPoSChain2026}. A transaction invokes a smart contract function all within one atomic unit: if any step reverts, the whole transaction rolls back and no state change persists~\cite{zhengOverviewSmartContracts2020}. Every step also costs \emph{gas}, paid in the network token POL by whoever sends the transaction~\cite{buterinNextgenerationSmartContract2014}. For Polymarket Exchange, that sender is always the official operator, so the operator bears the gas cost of a settlement even when the transaction reverts.

\paratitle{Transaction ordering} Unlike Ethereum~\cite{buterinNextgenerationSmartContract2014} where transactions usually route through private channels (\ie, PBS~\cite{ethereumProposerbuilderSeparation2026}), Polygon PoS exposes a public mempool in which a pending transaction is visible to anyone~\cite{seoevBiddingGamesReinforcement2025}. Validators order pending transactions largely by the gas fee each offers, so a party that wants its transaction to land ahead of another's can simply bid a higher fee~\cite{chenUnderstandingSecurityRisks2023}.

\paratitle{Settlement requirements} When \texttt{matchOrders} executes on-chain, it re-checks four conditions against current state: \textit{1)} the order's nonce must still match the owner's current nonce, advanced by a public \texttt{incrementNonce()} call that invalidates all of the owner's orders at once (V1 only); \textit{2)} the owner must still hold enough collateral or outcome tokens to deliver; \textit{3)} the allowance the owner granted the Exchange must still cover the transfer; and \textit{4)} the recipient of each ERC1155 transfer must accept the delivery callback~\cite{polymarketCtfexchange2026}.

\subsection{Ghost Fill Exploits: A Motivating Example}
\label{sec:bg:motivating-example}

\definecolor{cVuln}{RGB}{190,50,50}
\definecolor{cGreen}{HTML}{009E73}
\definecolor{cBlue}{HTML}{0072B2}
\definecolor{cRed}{RGB}{190,50,50}
\definecolor{cYellow}{HTML}{F0E442}

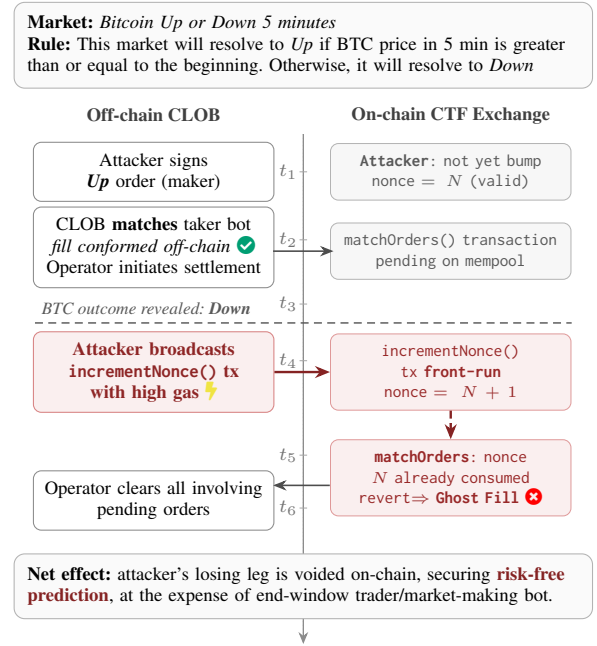
\begin{figure}[t]
\centering
\resizebox{\textwidth}{!}{%
\begin{tikzpicture}[
  A/.style    ={-{Stealth[length=1.8mm,width=1.4mm]}, line width=0.6pt,
                color=black!70},
  bad/.style  ={-{Stealth[length=2.0mm,width=1.6mm]}, line width=0.9pt,
                color=cVuln!70!black},
  taxis/.style={line width=0.5pt, color=black!40,
                -{Stealth[length=1.6mm,width=1.3mm]}},
  deadline/.style={densely dashed, line width=0.55pt, color=black!60},
  box/.style  ={rounded corners=3pt, line width=0.4pt, anchor=west,
                inner xsep=3pt, inner ysep=2pt, align=center,
                minimum height=0.8cm},
  evt/.style  ={box, draw=black!45, fill=white, font=\scriptsize},
  badevt/.style={box, draw=cVuln!60, fill=cVuln!8,
                 text=cVuln!75!black, font=\scriptsize\bfseries},
  status/.style={box, draw=black!30, fill=black!3, text=black!65,
                 font=\fontsize{6.5}{7.8}\selectfont\ttfamily},
  badstatus/.style={status, draw=cVuln!50, fill=cVuln!8,
                 text=cVuln!80!black},
  tlabel/.style={font=\scriptsize, text=black!55, fill=white, inner xsep=2pt, inner ysep=0.5pt},
  lane/.style ={font=\fontsize{6.8}{7.6}\selectfont\bfseries, align=center,
                text=black!70, rotate=90},
  dlbl/.style ={font=\fontsize{6.5}{7.8}\selectfont\itshape,
                text=black!65, inner sep=1.5pt},
  card/.style ={anchor=south west, rounded corners=3pt, inner xsep=7pt,
                inner ysep=3pt, font=\scriptsize, align=left,
                minimum height=1.12cm},
  scenecard/.style={card, fill=black!4, text=black!80,
                path picture={\fill[black!40]
                  (path picture bounding box.south west) rectangle
                  ([xshift=2.4pt]path picture bounding box.north west);}},
  effectcard/.style={card, fill=cVuln!6, text=black!80,
                path picture={\fill[cVuln!70]
                  (path picture bounding box.south west) rectangle
                  ([xshift=2.4pt]path picture bounding box.north west);}},
]

\def\yoff{0}
\def\yax{-0.72}
\def\yon{-1.44}
\def\xR{14.5}
\def\xD{6.95}

\node[lane] at (0.2,\yoff) {Off-chain\\CLOB};
\node[lane] at (0.2,\yon)  {On-chain\\Exchange};
\draw[taxis] (0.5,\yax) -- (\xR+0.1,\yax);

\node[evt, text width=2.45cm] (e1) at (0.55,\yoff)
  {Attacker signs an\\\textit{\textbf{Up}} order as maker};
\node[status, text width=2.45cm] (sPre) at (0.55,\yon)
  {\textbf{Attacker}: nonce $= N$\\(valid, not bumped)};

\node[evt, text width=3.1cm] (e2) at (3.45,\yoff)
  {CLOB \textbf{matches} a taker bot\\\textit{fill confirmed off-chain} \textcolor{cGreen}{\faCheckCircle}};
\node[status, text width=3.1cm] (eOp1) at (3.45,\yon)
  {Operator's \textbf{matchOrders()}\\pending in mempool};
\draw[A] ([xshift=0.5cm]e2.south) -- ([xshift=0.5cm]eOp1.north);

\draw[deadline] (\xD,0.47) -- (\xD,-1.9);
\node[dlbl, anchor=south] at (\xD,0.47)
  {BTC outcome revealed: \textit{\textbf{Down}}};

\node[badevt, text width=3.3cm] (eAtk) at (7.2,\yoff)
  {Attacker broadcasts high-gas\\\texttt{incrementNonce()} \textcolor{cYellow}{\faBolt}};
\node[badstatus, text width=3.3cm] (eBump) at (7.2,\yon)
  {\textbf{incrementNonce()} mined first\\nonce $= N+1$};
\draw[bad] ([xshift=0.5cm]eAtk.south) -- ([xshift=0.5cm]eBump.north);

\node[badstatus, text width=3.35cm] (sPost) at (10.95,\yon)
  {\textbf{matchOrders()}: nonce $N$ used\\\textsc{revert} $\Rightarrow$ \textbf{Ghost Fill} \textcolor{red}{\faTimesCircle}};
\draw[bad, densely dashed] (eBump.east) -- (sPost.west);

\node[evt, text width=3.35cm] (eOp) at (10.95,\yoff)
  {Operator clears all\\involved orders};
\draw[A] ([xshift=0.6cm]sPost.north) -- ([xshift=0.6cm]sPost.north |- eOp.south);

\foreach \x/\t in {1.8/{$t_1$}, 4.1/{$t_2$}, \xD/{$t_3$}, 8.0/{$t_4$}, 11.4/{$t_5$}, 14.0/{$t_6$}}
  \node[tlabel] at (\x,\yax) {\t};

\node[scenecard, text width=7.3cm] (scene) at (0.05,0.98)
  {{\bfseries Market:} \emph{Bitcoin Up or Down 5 minutes}.
  
   {\bfseries Rule:} This market will resolve to \emph{Up} if BTC price in 5 min is greater than or equal to the beginning. Otherwise, it will resolve to \emph{Down}.};
\node[effectcard, text width=6.2cm] (effect) at (8.1,0.98)
  {{\bfseries Net effect:} the attacker's losing leg is voided on-chain, securing a \textcolor{cVuln!75!black}{\textbf{risk-free prediction}} at the expense of the end-window trader or market-making bot.};

\end{tikzpicture}%
}
\caption{An example of \textit{Ghost Fill} exploits: In a 5-minute BTC price prediction market, an attacker exploits the time gap between the outcome of prediction and on-chain settlement by front-running a high gas \texttt{incrementNonce()} transaction to secure risk-free prediction.}
\Description{A horizontal timeline with an off-chain CLOB lane above and an on-chain exchange lane below. The attacker's order is matched off-chain, the outcome is revealed, the attacker front-runs the pending settlement with incrementNonce, and the settlement reverts as a Ghost Fill. Two cards on top describe the market and the net effect.}
\label{fig:motivating-nonce-bump}
\vspace{-1em}
\end{figure}

The settlement gap described in Section~\ref{sec:bg:polymarket} is not only a structural failure mode but also an opportunity for abuse. An attacker can deliberately mutate its own on-chain state inside the matching-to-settlement window to force the settlement to revert, thereby manufacturing a \textit{Ghost Fill} for profit. We call this behavior a \textit{Cancellation Attack}.

Figure~\ref{fig:motivating-nonce-bump} shows the attack on a 5-min Bitcoin market~\cite{polymarket5MinuteCryptoOdds2026}. The attacker signs an order in the last few seconds, betting that BTC will close \emph{Up}, and lets the CLOB match it against a counterparty, while the operator's \texttt{matchOrders} settlement remains pending in the mempool. Once the window closes with BTC \emph{Down}, the signed order has become a losing leg~\footnote{A \textit{leg} is one of the orders that compose a matched trade; a losing leg here is such an order whose predicted outcome turns out to be wrong.}. The attacker then front-runs the pending settlement with a high-gas \texttt{incrementNonce()} call. Because this call advances its nonce, the stale order now fails the on-chain nonce check and \texttt{matchOrders} reverts. The attacker thus secures a risk-free prediction at the expense of the counterparty end-window trader or market-making bot.

\begin{noticebox}
\textbf{Definitions:} A \textit{Ghost Fill} is a user-side phenomenon: CLOB reports an off-chain fill to both sides, but the on-chain settlement transaction reverts, leaving no actual transfers; A \textit{Cancellation Attack} is an attacker-side behavior: a deliberate action taken after the off-chain match and before settlement that invalidates the matched order so its settlement is forced to revert.
\end{noticebox}

\section{Threat Model}
\label{sec:threat-model}

A trade in Polymarket involves several roles.
The \emph{owner} of an order holds the collateral or outcome tokens that back it; in a match, the owner of the crossing order is the \emph{taker} and the owners of the resting orders are the \emph{makers} (Section~\ref{sec:bg:polymarket}).
The \emph{signer} signs the order, either as the owner itself or as a delegate key acting on its behalf.
The \emph{operator} is the Polymarket-controlled account that submits the on-chain \texttt{matchOrders} transaction once the off-chain order book reports a match; Polymarket runs several operator accounts in parallel to sustain throughput.
The \emph{attacker} is an ordinary market participant who controls one or more funded accounts, either externally owned accounts (EOAs) or contract accounts (CAs)~\cite{buterinNextgenerationSmartContract2014}. Its \emph{companions} are further attacker-controlled accounts used to collect profit.

The attacker has only the capabilities of a regular Polymarket user. Between the moment the CLOB reports a match and the moment the operator's \texttt{matchOrders} settles on-chain, it can act on the accounts it controls to make the pending settlement revert, outbidding the settlement with a higher gas price when necessary~\cite{maSurvivingDarkForest2025}.
Since Polymarket may blacklist abusive addresses, it can also operate many unlinkable accounts as a Sybil cluster~\cite{douceurSybilAttack2002}.

This capability gives the attacker a free option on every match it joins: it keeps the benefit of a fill that turns out favorable and voids one that would lock in a loss. Repeated at scale, the same cancellations also exhaust the operator's gas and settlement throughput, amounting to a denial of service against the platform.

\paratitle{Scope} Our measurements span two protocol generations and six contracts of Polymarket (Table~\ref{tab:contracts}). In \emph{V1}, an on-chain \texttt{matchOrders} call is sent to one of two Fee Module contracts for binary and neg-risk markets. Each Fee Module wraps an underlying Exchange that performs the token transfers. V1 uses USDC.e as collateral and gives each order a per-owner \texttt{nonce}, incremented through a public \texttt{incrementNonce()} function.
The platform-wide cutover to \emph{V2} occurred on \vTwoCutover. V2 routes \texttt{matchOrders} directly to two new exchange contracts and removes the Fee Modules entirely~\cite{polymarketMigratingCLOBV22026}. It switches collateral to pUSD and replaces the nonce with a timestamp-based uniqueness check. On May 4, 2026, Polymarket further moved participants to Deposit Wallets (Section~\ref{sec:disc:mitigation}). Our study window, \studyStart{} to \studyEnd, covers both upgrades.
\section{\tool}
\label{sec:methodology}

\definecolor{cVuln}{RGB}{190,50,50}
\definecolor{cGreen}{HTML}{009E73}
\definecolor{cBlue}{HTML}{0072B2}
\definecolor{cOrange}{HTML}{E69F00}

\begin{figure}[t]
\centering
\resizebox{\textwidth}{!}{%
\begin{tikzpicture}[
  A/.style     ={-{Stealth[length=2.6mm,width=2.2mm]}, line width=1.2pt,
                 color=black!72},
  Aeng/.style  ={-{Stealth[length=2.2mm,width=1.8mm]}, line width=0.9pt,
                 color=black!75},
  Arule/.style ={-{Stealth[length=1.6mm,width=1.3mm]}, line width=0.6pt,
                 color=cVuln!70!black},
  src/.style   ={rounded corners=2.5pt, draw=black!50, line width=0.5pt,
                 fill=white, inner xsep=3pt, inner ysep=2pt,
                 font=\scriptsize\bfseries, align=center,
                 text width=2.2cm, minimum height=0.5cm},
  mod/.style   ={rounded corners=3.5pt, draw=black!55, line width=0.6pt,
                 fill=white, inner xsep=4pt, inner ysep=4pt,
                 font=\small\bfseries, align=center, minimum height=1.85cm},
  rmod/.style  ={mod, minimum height=1.6cm},
  modhdr/.style={font=\small\bfseries, text=black!85, align=center},
  detail/.style={font=\fontsize{7.0}{8.2}\selectfont\ttfamily,
                 text=black!75, align=left, inner sep=0pt},
  fact/.style  ={font=\fontsize{7.0}{8.2}\selectfont\ttfamily, text=black!80,
                 anchor=west, inner xsep=2.5pt, inner ysep=1.6pt,
                 text width=1.68cm, align=left, rounded corners=1.5pt},
  rbox/.style  ={rounded corners=2.5pt, draw=cVuln!55, line width=0.55pt,
                 fill=cVuln!6, font=\fontsize{6.2}{7.6}\selectfont\ttfamily,
                 text=cVuln!82!black, align=center, text width=1.55cm,
                 minimum height=0.44cm, inner ysep=2pt},
  rq/.style    ={rounded corners=3pt, draw=cBlue!55, line width=0.7pt,
                 fill=cBlue!7, inner xsep=3pt, inner ysep=2pt,
                 align=center, text width=1.8cm, minimum height=1.0cm,
                 font=\fontsize{7.2}{8.0}\selectfont},
  hdr/.style   ={font=\fontsize{8.0}{8.8}\sffamily\bfseries, text=black!60,
                 align=center},
  groupbox/.style={draw=black!30, dashed, line width=0.45pt,
                 rounded corners=5pt, fill=white,
                 inner xsep=4pt, inner ysep=6pt},
  enginebox/.style={draw=black!45, line width=0.6pt,
                 rounded corners=5pt, fill=black!2,
                 inner xsep=5pt, inner ysep=6pt},
]

\def\xsrc{0}
\def\xcol{2.9}
\def\xtr{5.66}
\def\xfb{8.71}
\def\xru{12.385}
\def\xout{16.4}
\def\yeng{0}
\def\yreuse{-2.75}

\node[src] (onchain)      at (\xsrc, 0.62) {On-chain reverts};
\node[src] (onchaintrace) at (\xsrc, 0.0)  {On-chain traces};
\node[src] (offchain)     at (\xsrc,-0.62) {Off-chain markets};
\node[src, minimum height=0.8cm] (crosschain) at (\xsrc,\yreuse) {Cross-chain\\Verified Contracts};

\begin{scope}[on background layer]
  \node[groupbox, fit=(onchain)(onchaintrace)(offchain)(crosschain),
    label={[hdr, anchor=south, yshift=-1pt]north:Data Source}] (feedbox) {};
\end{scope}
\coordinate (mdB) at ($(offchain.south)!0.5!(crosschain.north)$);
\draw[line width=0.5pt, color=black!35]
  ([xshift=6pt]feedbox.west |- mdB) -- ([xshift=-6pt]feedbox.east |- mdB);

\node[mod, text width=1.45cm] (collector) at (\xcol,\yeng) {Tx\\Collector};

\node[mod, text width=2.7cm] (tracer) at (\xtr,\yeng) {};
\node[modhdr, anchor=north] at ([yshift=-4pt]tracer.north) {Tx Tracer};
\node[detail, anchor=west] (d1) at ([xshift=5pt,yshift=-0.05cm]tracer.west) {matchOrders(\ldots)};
\node[detail, anchor=north west] (d2) at ([xshift=9pt,yshift=-3pt]d1.south west) {transferFrom};
\node[detail, anchor=north west, fill=cVuln!12, inner xsep=2pt, inner ysep=1.5pt] (d3)
  at ([xshift=9pt,yshift=-3pt]d2.south west)
  {\textcolor{cVuln!80!black}{\textbf{REVERT}}\ 0x\ldots};
\draw[line width=0.4pt, color=black!45]
  ([xshift=3pt]d1.south west) |- (d2.west);
\draw[line width=0.4pt, color=black!45]
  ([xshift=3pt]d2.south west) |- (d3.west);

\node[mod, text width=2.0cm] (facts) at (\xfb,\yeng) {};
\node[modhdr, anchor=north] at ([yshift=-4pt]facts.north) {Fact Builder};
\node[fact, fill=black!7] (f1) at ([xshift=6pt,yshift=0.07cm]facts.west) {order(t,a,n)};
\node[fact, fill=cBlue!11, anchor=north west] (f2) at ([yshift=-2.2pt]f1.south west) {err(c,$\varepsilon$)};
\node[fact, fill=cOrange!18, anchor=north west] (f3) at ([yshift=-2.2pt]f2.south west) {cross(a,k,t,d)};

\node[mod, text width=3.95cm] (rules) at (\xru,\yeng) {};
\node[modhdr, anchor=north] at ([yshift=-4pt]rules.north) {Rule Matching};
\node[rbox] (rbnonce)   at (\xru-1.0, 0.0)  {proxy\_trap};
\node[rbox] (rbproxy)   at (\xru+1.0, 0.0)  {nonce\_bump};
\node[rbox] (rbapprove) at (\xru-1.0,-0.6)  {approve\_revoke};
\node[rbox] (rbbalance) at (\xru+1.0,-0.6)  {balance\_drain};
\draw[Arule] (rbnonce.east)   -- (rbproxy.west);
\draw[Arule] (rbproxy.south)  -- (rbbalance.north);
\draw[Arule] (rbbalance.west) -- (rbapprove.east);

\draw[Aeng] (collector.east) -- (tracer.west);
\draw[Aeng] (tracer.east)    -- (facts.west);
\draw[Aeng] (facts.east)     -- (rules.west);

\begin{scope}[on background layer]
  \node[enginebox, fit=(collector)(tracer)(facts)(rules),
    label={[hdr, anchor=south, yshift=-1pt]north:Tx Analysis Engine}] (engbox) {};
\end{scope}

\node[rmod, text width=1.45cm] (selx) at (\xcol,\yreuse) {};
\node[modhdr, anchor=north] at ([yshift=-4pt]selx.north) {Selector\\[-2pt]Extraction};
\node[detail, align=center, anchor=south] at ([yshift=5pt]selx.south) {4-byte set};

\node[rmod, text width=5.1cm] (jac) at ($(\xtr,\yreuse)!0.5!(\xfb,\yreuse)+(-0.15,0)$) {};
\node[modhdr, anchor=north] at ([yshift=-4pt]jac.north) {Jaccard Scoring};
\coordinate (bc) at ([xshift=0.55cm,yshift=-0.22cm]jac.west);
\def\bcs{0.21}
\begin{scope}[shift={(bc)}]
  \node[anchor=east, font=\fontsize{6.6}{7}\selectfont\sffamily\bfseries, text=black!60, inner sep=1pt] at (-0.08, 0.17) {A};
  \node[anchor=east, font=\fontsize{6.6}{7}\selectfont\sffamily\bfseries, text=black!60, inner sep=1pt] at (-0.08,-0.17) {B};
  \draw[cBlue!80!black, line width=1.0pt] (0,0.17) -- (7*\bcs,0.17);
  \foreach \i in {0,...,7} \draw[cBlue!80!black, line width=1.1pt] (\i*\bcs,0.08) -- (\i*\bcs,0.26);
  \draw[cBlue!80!black, line width=1.0pt] (0,-0.17) -- (4*\bcs,-0.17);
  \draw[cBlue!80!black, line width=1.0pt] (6*\bcs,-0.17) -- (7*\bcs,-0.17);
  \foreach \i in {0,1,2,3,4,6,7} \draw[cBlue!80!black, line width=1.1pt] (\i*\bcs,-0.26) -- (\i*\bcs,-0.08);
  \draw[cVuln!75, line width=0.6pt] (5*\bcs-0.06,-0.23) -- (5*\bcs+0.06,-0.11);
  \draw[cVuln!75, line width=0.6pt] (5*\bcs-0.06,-0.11) -- (5*\bcs+0.06,-0.23);
  \foreach \i in {0,1} \draw[cOrange, line width=1.0pt] (7*\bcs+0.22+\i*0.09,-0.25) -- (7*\bcs+0.22+\i*0.09,-0.09);
\end{scope}
\node[anchor=west, font=\scriptsize, text=black!80] at ([xshift=2.35cm,yshift=-0.22cm]jac.west)
  {$J(A,B)=\dfrac{|A\cap B|}{|A\cup B|}$};

\node[rmod, text width=3.95cm] (exam) at (\xru,\yreuse) {};
\node[modhdr, anchor=north] at ([yshift=-4pt]exam.north) {Manual Examination};
\node[detail, align=center, anchor=south] at ([yshift=5pt]exam.south)
  {$J\geq0.5$: 494 candidates\\$\rightarrow$ \reusedContracts reuses, \reuseChainCount chains};

\draw[Aeng] (selx.east) -- (jac.west);
\draw[Aeng] (jac.east)  -- (exam.west);

\begin{scope}[on background layer]
  \node[enginebox, fit=(selx)(jac)(exam),
    label={[hdr, anchor=south, yshift=-1pt]north:Reuse Analysis}] (reubox) {};
\end{scope}

\newcommand{\rqnum}[1]{{\scriptsize\bfseries\textcolor{cBlue!75!black}{#1}}}
\newcommand{\rqcapfont}{\fontsize{7.2}{8.4}\selectfont\color{black!72}}
\node[rq] (rq1) at (\xout, 0.58)
  {\rqnum{RQ1}\\\rqcapfont Ghost Fill\\prevalence};
\node[rq] (rq2) at (\xout,-0.58)
  {\rqnum{RQ2}\\\rqcapfont Cancellation\\Attacks};
\node[rq] (rq3) at (\xout,\yreuse)
  {\rqnum{RQ3}\\\rqcapfont Risk spread};

\begin{scope}[on background layer]
  \node[groupbox, fit=(rq1)(rq2)(rq3),
    label={[hdr, anchor=south, yshift=-1pt]north:Results}] (outbox) {};
\end{scope}

\draw[A] (feedbox.east |- engbox.west) -- (engbox.west);
\draw[A] (engbox.east |- rq1) -- (rq1.west);
\draw[A] (engbox.east |- rq2) -- (rq2.west);
\draw[A] (crosschain.east) -- (reubox.west |- crosschain.east);
\draw[A] (reubox.east |- rq3) -- (rq3.west);

\end{tikzpicture}%
}
\caption{Overview of \tool Pipeline.}
\Description{GhostHunter pipeline. On-chain reverts, traces, and off-chain market data feed a transaction analysis engine (collector, tracer, fact builder, and rule matching) that answers RQ1 and RQ2; cross-chain verified contracts feed a reuse analysis (selector extraction, Jaccard scoring against the Polymarket contracts, and manual examination) that answers RQ3.}
\label{fig:architecture}
\vspace{-1em}
\end{figure}
Figure~\ref{fig:architecture} presents the overview of \tool. 
We first collect all reverted on-chain \texttt{matchOrders} transactions of Polymarket and trace their on-chain execution to recover the revert reason and classify each failure.
This step supports the measurement of \textit{Ghost Fill} prevalence and impact (RQ1).
\tool then performs causal attribution for attacker-caused cases. It turns each reverted transaction, together with the participant actions before it, into relational facts. Vector-specific rules over these facts then emit a structured record containing the attack vector, the causal action, and the addresses involved (RQ2).
Finally, \tool extracts the function-selector set of every verified contract on 401 chains, scores it against the official Polymarket contracts, and manually examines the closest matches to identify deployments that reuse Polymarket-like designs (RQ3).

\subsection{Data Collection}
\label{sec:method:data}
We conduct our measurement from \studyStart, when the V1 Fee Module went live~\cite{polymarketMigratingCLOBV22026}, to \studyEnd, spanning both the V1-to-V2 cutover and the Deposit Wallet upgrade (see Section~\ref{sec:disc:mitigation}). We draw on three data sources.

\paratitle{On-Chain Transaction} 
Our primary dataset consists of all reverted \texttt{matchOrders} transactions sent to the settlement contracts listed in Table~\ref{tab:contracts}. We query Google BigQuery's public Polygon tables~\cite{googlebigqueryPolygonBlockchainDataset2026} for transactions to these contracts whose receipts indicate failure, recording their block numbers, contract addresses, transaction hashes, timestamps, transaction indices, raw calldata, gas usage, and gas prices. We further retrieve the full execution trace of each reverted transaction from an Alchemy archive node~\cite{alchemyAlchemyBlockchainInfrastructure2026}. This process yields \ghostFillCount reverted transactions.

\paratitle{Off-Chain Market} 
On-chain data identifies settlement contracts and outcome tokens, but not the markets associated with those tokens.
We therefore supplement the transaction dataset with Polymarket's Gamma API~\cite{polymarketGammaAPI2026}, which maps each outcome token to its market slug~\cite{polymarketMarketsEvents2026}, event category, resolution time, and liquidity reward parameters.
For the avoided-loss analysis in Section~\ref{sec:disc:weaponization}, we also record the resolution outcome of each market.

\paratitle{Cross-Chain Verified Contracts} 
To assess whether Polymarket-like designs are reused beyond Polymarket, we use the verified-source dataset published by Sourcify~\cite{sourcifySourcifySupportedChains2026}, comprising 32{,}103{,}371 contracts across 401 chains. We query and filter this dataset through BigQuery to identify contracts with interfaces similar to Polymarket's exchange contracts. All SQL queries list in Appendix~\ref{app:sql}.

\subsection{Transaction Analysis}
\label{sec:method:design}

\tool analyzes each reverted \texttt{matchOrders} transaction by turning raw on-chain evidence into a uniform set of facts, and then applying attribution rules over those facts. For each transaction, \tool first decodes the \texttt{matchOrders} calldata to recover the orders, participants, token amounts, and settlement contract. It then parses the execution trace to identify the internal calls made during settlement, the call that failed, and the corresponding revert reason. We further map each revert reason to its \texttt{require}/\texttt{revert} site in the official source code, which allows us to group reverts into the failure surfaces.

\paratitle{Evidence Schema} 
The evidence needed to explain a revert may come from multiple sources. Some live in the failed transaction itself: the decoded order fields and the execution trace. Confirming whether the revert was deliberate also requires the causal transaction that changed the participant's state before settlement. We represent this evidence as a flat set of relational facts, summarized in Figure~\ref{fig:facts}. We organize the facts into three groups:
\textit{skeleton facts} describe the order and its settlement context; \textit{frame facts }describe calls inside the failed transaction; \textit{causal facts} describe participant actions that precede the settlement.

\begin{figure}[t]
    \centering
    {\footnotesize
    \setlength{\arraycolsep}{4pt}
    \renewcommand{\arraystretch}{0.95}
    \begin{minipage}[t]{0.49\textwidth}
    \centering
    $
        \begin{array}[t]{ll}
        \multicolumn{2}{l}{\textbf{Skeleton} \text{ (decoded order and context)}} \\
          {revert}(t) & \text{reverted}\ \texttt{matchOrders} \\
          {exchange}(t, x) & \text{settlement exchange}\ x \\
          {maker}(t, m),\ {taker}(t, m) & m\ \text{owns a maker / taker order} \\
          {order}(t, a, n) & a\text{'s order carries nonce}\ n \\
          {signed}(t, a, s) & a\ \text{signed its order at}\ s \\[2pt]
        \multicolumn{2}{l}{\textbf{Frame} \text{ (execution trace of }t\text{)}} \\
          {frame}(t, c) & \text{internal call}\ c \\
          {sel}(c, \sigma) & \text{selector of}\ c \\
          {target}(c, w) & \text{callee}\ w\ \text{of}\ c \\
          {err}(c, \varepsilon) & c\ \text{reverted with}\ \varepsilon \\
          {failed}(c) & c\ \text{ended in any error}
        \end{array}
    $
    \end{minipage}\hfill
    \begin{minipage}[t]{0.49\textwidth}
    \centering
    $
        \begin{array}[t]{ll}
        \multicolumn{2}{l}{\textbf{Frame} \text{ (continue.)}} \\
          {holder}(c, a) & \text{owner}\ a\ \text{of failed transfer} \\
          {need}(c, k) & a\ \text{owes}\ k\ \text{across all legs of}\ t \\[2pt]
        \multicolumn{2}{l}{\textbf{Causal} \text{ (actions before settlement)}} \\
          {bump}(a, x, n, t, d) & d\ \text{raised}\ a\text{'s nonce on}\ x\ \text{past}\ n \\
          {cross}(a, k, t, d) & d\ \text{last took}\ a\text{'s balance below}\ k \\
          {approve}(a, x, t, d, v) & d\ \text{last set}\ x\text{'s allowance to}\ v \\[2pt]
        \multicolumn{2}{l}{\textbf{Derived}} \\
          {part}(t, a) & a\ \text{is a participant of}\ t \\
          {path}(d, a) & a\ \text{or its signer is on}\ d\text{'s call path} \\
          {settle}(d) & d\ \text{is a settlement transaction} \\
          {after}(d, s) & d\ \text{landed after time}\ s
        \end{array}
    $
    \end{minipage}
    }
    \caption{Relational fact schema underlying \tool's rules. Every causal transaction $d$ precedes $t$ by block and transaction index, and ``last'' refers to the final such transaction.}
    \Description{Four groups of relational facts: skeleton facts about the reverted transaction, frame facts from its execution trace, causal facts about actions before settlement, and derived facts.}
    \label{fig:facts}
    \vspace{-1em}

\end{figure}

\paratitle{Rule Matching} 
Each rule describes one attack vector as a conjunction of facts. The rule first checks whether the revert is consistent with the attack vector. For example, a nonce-bump rule requires an \texttt{InvalidNonce} revert, while an allowance-revoke rule requires a failed \texttt{transferFrom} caused by insufficient allowance. If this holds, the rule then looks for causal evidence showing that a participant-controlled action occurred before settlement and explains the failure. The revert reason narrows the plausible explanations, and the causal evidence decides whether the revert is an attack. Rules are applied in a fixed priority order (Algorithm~\ref{alg:classify}).

\paratitle{Rule Construction}
Because there is no ground-truth dataset of \textit{Cancellation Attacks}, we build the rule set through iterative snowball sampling~\cite{goodmanSnowballSampling1961,wohlinEmpiricalResearchMethods2003}. We first group reverted transactions by revert reason and manually inspect samples from each group, including their calldata, execution traces, surrounding transactions, and state changes. 
When we observe a recurring mechanism, we encode it as a conjunction of on-chain facts and add it to the rule set. We then rerun \tool, audit the matched cases for false positives, and sample the remaining unmatched cases to identify mechanisms missed by the current rule set. This process repeats until additional samples yield no new patterns.
The resulting rules serve as our definition for attack-caused \textit{Ghost Fills}: a revert is attributed to an attack when a participant-controlled action invalidates an already-matched order before its settlement confirms.

\subsection{Attack Vector Rule}
\label{sec:method:rules}
We find four attack vectors and encode each as a rule:

\paratitle{Proxy Trap} 
This occurs when a participant's wallet refuses the outcome tokens that settlement delivers to it. As Section~\ref{sec:bg:polymarket} describes, an ERC1155 recipient contract can reject a delivery through its \texttt{onERC1155Received} hook, and a rejected delivery reverts the whole settlement. Most trap wallets are Gnosis Safes, which route this hook to their fallback handler. In the dominant variant, the attacker keeps the official Safe proxy and singleton and replaces only the fallback handler with a contract whose hook always fails. The wallet's code is identical to that of an ordinary Safe, yet every delivery to it reverts. Unlike the other vectors, the attack needs no causal transaction per settlement. The trap is set once and fires on every match that involves the wallet, so the rule inspects only the failed transaction itself. It matches any revert in which an \texttt{onERC1155Received} callback to a participant wallet fails:

\vspace{0.5em}
\begin{MintedVerbatim}[commandchars=\\\{\}]
\PYG{n+nf}{proxy\PYGZus{}trap}(T) \PYG{k}{:-}
    \PYG{n+nf}{revert}(T), \PYG{n+nf}{frame}(T, C), \PYG{n+nf}{sel}(C, \PYG{l+s+sAtom}{onERC1155Received}),
    \PYG{n+nf}{target}(C, W), \PYG{n+nf}{failed}(C), \PYG{n+nf}{part}(T, W).
\end{MintedVerbatim}
\vspace{0.5em}

The fact $\mathit{failed}(C)$ covers every way the callback can fail, whether a direct revert, an out-of-gas error from deliberate gas burning, or a stack overflow from forced recursion.

\paratitle{Nonce Bump}
This occurs when a participant invalidates an already matched order by increasing its own nonce on the exchange~\cite{polymarketCtfexchange2026}. The settlement then fails with \texttt{InvalidNonce} because the nonce embedded in the signed order no longer matches. The rule looks for the transaction that made the order's nonce stale:

\vspace{0.5em}
\begin{MintedVerbatim}[commandchars=\\\{\}]
\PYG{n+nf}{nonce\PYGZus{}bump}(T) \PYG{k}{:-}
    \PYG{n+nf}{revert}(T), \PYG{n+nf}{err}(_, \PYG{l+s+sAtom}{InvalidNonce}),
    \PYG{n+nf}{order}(T, A, N), \PYG{n+nf}{exchange}(T, X), \PYG{n+nf}{bump}(A, X, N, T, D).
\end{MintedVerbatim}
\vspace{0.5em}

The V1 exchange accepts an order only when its nonce $N$ equals the current nonce of its owner $A$, and each \texttt{incrementNonce()} call raises the caller's nonce by one. An order therefore goes stale at exactly one transaction, the one that moved $A$'s nonce from $N$ to $N{+}1$. The rule takes the first order in the reverted transaction that fails validation and uses the exchange's public \texttt{nonces} getter to find the first block in which $A$'s nonce exceeds $N$. Within that block, \tool keeps the transaction $D$ whose trace shows a successful \texttt{incrementNonce()} call from $A$ to the exchange, and $D$ must precede the revert. Because the exchange keys each nonce by its caller, only $A$ can make this call. The rule accepts it whether it comes from a top-level transaction, a Gnosis Safe (including a \texttt{multiSend} batch), Polymarket's proxy-wallet factory, or a smart-account wallet.

\paratitle{Allowance Revoke}
This occurs when a participant revokes or reduces the exchange's allowance on the collateral token after signing the order, causing \texttt{transferFrom} to fail with insufficient allowance:

\vspace{0.5em}
\begin{MintedVerbatim}[commandchars=\\\{\}]
\PYG{n+nf}{allowance\PYGZus{}revoke}(T) \PYG{k}{:-}
    \PYG{n+nf}{revert}(T), \PYG{n+nf}{frame}(T, C), \PYG{n+nf}{sel}(C, \PYG{l+s+sAtom}{transferFrom}),
    \PYG{n+nf}{err}(C, \PYG{l+s+sAtom}{InsufficientAllowance}), \PYG{n+nf}{holder}(C, A), \PYG{n+nf}{need}(C, K),
    \PYG{n+nf}{exchange}(T, X), \PYG{n+nf}{approve}(A, X, T, D, V), V < K,
    \PYG{n+nf}{signed}(T, A, S), \PYG{n+nf}{after}(D, S).
\end{MintedVerbatim}
\vspace{0.5em}

Our rule identifies the account $A$ whose \texttt{transferFrom} failed and the amount $K$ that the settlement needs from it, summed over every leg $A$ pays in the transaction. It then reads the collateral token's \texttt{Approval} events with $A$ as the owner and the exchange as the spender, which captures allowance changes made through both EOAs and proxy wallets. The rule takes the last allowance write $D$ before the revert, ordered by block and transaction index, and labels the revert only when $D$ set the allowance below $K$ and landed after the order was signed.

\paratitle{Balance Drain} 
This occurs when a participant moves collateral out of the funding wallet before settlement, leaving the exchange's \texttt{transferFrom} with insufficient balance. The rule looks for the transaction that pushed the balance below what settlement needs:

\vspace{0.5em}
\begin{MintedVerbatim}[commandchars=\\\{\}]
\PYG{n+nf}{balance\PYGZus{}drain}(T) \PYG{k}{:-}
    \PYG{n+nf}{revert}(T), \PYG{n+nf}{frame}(T, C), \PYG{n+nf}{sel}(C, \PYG{l+s+sAtom}{transferFrom}),
    \PYG{n+nf}{err}(C, \PYG{l+s+sAtom}{InsufficientBalance}), \PYG{n+nf}{holder}(C, A), \PYG{n+nf}{need}(C, K),
    \PYG{n+nf}{cross}(A, K, T, D), \PYG{k}{not} \PYG{n+nf}{settle}(D),
    \PYG{n+nf}{path}(D, A), \PYG{n+nf}{signed}(T, A, S), \PYG{n+nf}{after}(D, S).
\end{MintedVerbatim}
\vspace{0.5em}

\todo{after(D,S) only applies to V2 orders (V1 orders carry no signing time). Allowance Revoke says ``for V2 orders'' but Balance Drain does not; make the V1/V2 handling consistent in both rules and Datalog.}
As with Allowance Revoke, the rule starts from the holder $A$ and the required amount $K$. It rebuilds $A$'s collateral balance from the token's transfer events, ordered by block and transaction index, and finds the last transaction $D$ before the revert that took the balance from at least $K$ to below $K$. The search walks backward over growing block ranges and stops at the first such crossing.
Three checks then decide whether $D$ is a drain. First, $D$ must not be another settlement, since a crossing caused by a concurrent fill is a settlement race. Second, $A$ or the signer of its order must appear on $D$'s call path, whether $D$ is a direct transfer, a Gnosis Safe \texttt{execTransaction}, or a relayed meta-transaction. Third, $D$ must land after $A$ signed the order.

When several rules match the same revert, \tool keeps the label of the first in the order above: \texttt{proxy\_trap}, \texttt{nonce\_bump}, \texttt{allowance\_revoke}, and \texttt{balance\_drain}. Specific failure mechanisms thus take precedence over the more general balance-drain rule.

\begin{figure}[t]
  \centering
  \includegraphics[width=\textwidth]{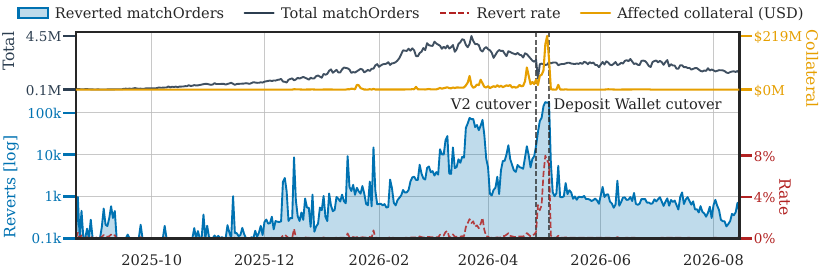}
  \caption{Overview of daily \texttt{matchOrders} transactions.}
  \Description{Daily time series from August 2025 to August 2026 of reverted matchOrders (log scale), total matchOrders, the daily revert rate, and affected collateral, with the V2 cutover and the Deposit Wallet cutover marked.}
  \label{fig:rq1-reverts-daily}
  \vspace{-1.5em}

\end{figure}

\subsection{Cross-Chain Reuse Analysis}
\label{sec:method:spread}
To assess how widely the flawed exchange design has spread beyond Polymarket, we search for contracts whose on-chain interface resembles the official Polymarket exchange contracts. We base this search on each contract's function-selector set.

\paratitle{Selector-Set Similarity} 
Each externally callable function is identified on-chain by a \emph{function selector} (\ie, the first four bytes of the Keccak hash of its signature~\cite{chenSigRecAutomaticRecovery2022}.)
The selector set of a contract thus serves as a fingerprint for identifying code reuse.
We quantify the similarity between two contracts using the Jaccard index~\cite{fletcherComparingSetsPatterns2018}: for two selector sets $A$ and $B$, $J(A,B) = \frac{|A \cap B|}{|A \cup B|}$.
We take the four official Polymarket V1 contracts in Table~\ref{tab:contracts} as search targets and recover their selector sets from the verified source code in Sourcify. We then compare every other Sourcify-verified contract against them using the Jaccard score, retaining the maximum score for each contract. 

\paratitle{Examination} 
The selector scan produces a ranked list of candidate contracts, which we narrow through several filtering steps. First, we remove testnet deployments and contracts with no evidence of real use. For each remaining contract, we collect its transaction count and check whether it was still active at the end of our study window. 
Two authors then independently inspect the remaining candidates. They remove contracts whose selector overlap is coincidental and read each candidate's verified source to confirm that it implements a Polymarket-like hybrid exchange design.
Finally, for confirmed deployments, we identify the operating entity using block-explorer labels~\cite{polygonscanPolygonPoSChain2026} and public web records, and obtain total value locked from DeFiLlama~\cite{defillamaDefillama2026} when available~\cite{chenUnderstandingEthereumGraph2018}.
This yields \reusedContracts contracts across \reuseChainCount chains, which we analyze in Section~\ref{sec:rq3}.

\section{Answer to RQ1: Prevalence of Ghost Fills}
\label{sec:rq1}

Over the study window, \tool finds \ghostFillCount reverted \texttt{matchOrders} transactions, each a \textit{Ghost Fill}. We measure their volume over time (Section~\ref{sec:rq1:volume}), failure surfaces (Section~\ref{sec:rq1:surface}), affected markets and users (Section~\ref{sec:rq1:affected}), and financial impact (Section~\ref{sec:rq1:impact}).

\subsection{Overall Volume and Temporal Distribution}
\label{sec:rq1:volume}

\textit{Ghost Fills} are widespread. Across the window, \tool labels \ghostFillCount reverted settlement transactions involving \ghostFillParticipants distinct participants. The two protocol versions contribute similar totals, 1{,}090{,}017 under V1 and 947{,}287 under V2, on very different clocks. V1 accumulated its volume over roughly nine months, while V2 reached 862{,}423 in its first week at comparable daily settlement volume.

\finding{V2's first week produced nearly as many \textit{Ghost Fills} (862{,}423) as V1's full nine months (1{,}090{,}017). At comparable daily settlement volume, V2's daily revert rate runs more than an order of magnitude above V1's.}

The revert rate rose sharply over time (Figure~\ref{fig:rq1-reverts-daily}). Through late 2025, reverts were sporadic, typically a few hundred per day against a rising settlement load. From early 2026, they increased by more than two orders of magnitude; on the worst day, \peakDailyRate of all \texttt{matchOrders} transactions reverted. The worst hour came on May 4, when 23{,}596 of 97{,}048 transactions (\peakHourlyRate) reverted between 11:00 and 12:00 UTC. The Deposit Wallet upgrade on May 4, 2026 cut the daily rate from roughly 8\% to 0.3\% within two days. For the rest of the window, the rate averaged 0.04\% and never exceeded 0.25\%.

\subsection{A Taxonomy of Failure Surfaces}
\label{sec:rq1:surface}

A \texttt{matchOrders} can fail at several checks as depicted in Section~\ref{sec:bg:settlement}. We decode each transaction's revert reason, map it to the corresponding \texttt{require} or \texttt{revert} site in the official exchange source (Section~\ref{sec:method:design}), and group the sites into the eight failure surfaces in Table~\ref{tab:rq1-failure-surface}. Each revert reason maps to one surface, and the taxonomy covers 99.9\% of reverts. A trap wallet can make its callback fail with a reason borrowed from another check (Section~\ref{sec:rq2:proxy-trap}), so we assign every revert whose ERC1155 callback frame failed to the callback surface, whatever reason it reports.

\begin{table}[t]
  \centering
  \caption{Failure classification by failure surface.}
  \vspace{-1em}
  \label{tab:rq1-failure-surface}
  \footnotesize
  \setlength{\tabcolsep}{4pt}
  \begin{tabular}{l r r r r}
    \toprule
    Failure surface & V1 & V2 & Total & \% \\
    \midrule
    Insufficient Balance                      & 946{,}669   & 123{,}104 & 1{,}069{,}773 & 52.5 \\
    ERC1155 callback             & 56{,}122    & 748{,}119 & 804{,}241   & 39.5 \\
    Insufficient Allowance                    & 21{,}259    & 49{,}292  & 70{,}551    & 3.5 \\
    Order validation (operator)  & 39{,}184    & 4{,}815   & 43{,}999    & 2.2 \\
    Order validation (user)      & 25{,}790    & 0         & 25{,}790    & 1.3 \\
    Fee computation              & 0           & 21{,}076  & 21{,}076    & 1.0 \\
    Condition token              & 265         & 538       & 803          & $<0.1$ \\
    Other / unknown              & 728         & 343       & 1{,}071     & $<0.1$ \\
    \midrule
    Total                        & 1{,}090{,}017 & 947{,}287 & 2{,}037{,}304 & 100 \\
    \bottomrule
  \end{tabular}
  \vspace{-1em}

\end{table}

Two surfaces account for the majority. Insufficient balance failures make up 52.5\% of all reverts: the exchange's \texttt{transferFrom} found a short balance when settlement executed.
A reverted ERC1155 receiver callback explains another 39.5\%: the exchange could not deliver outcome tokens because the recipient wallet rejected the callback.
An insufficient allowance (3.5\%) blocks the same transfer one step earlier. Operator-side order validation (2.2\%) covers reverts that operator itself could have caught, such as an expired or already-filled order.
User-side order validation (1.3\%) is \texttt{InvalidNonce} bucket, and on-chain fee computation (1.0\%) is the V2-only \texttt{FeeExceedsMaxRate} fault.

The version split shows that the dominant failure surface changed after the cutover. Under V1, reverts concentrate in Balance (946{,}669), and all \texttt{InvalidNonce} failures (25{,}790) occur before V2 removed the nonce-based vector. Under V2, reverts concentrate in the ERC1155 callback surface (748{,}119 of 947{,}287), and the V2-only fee-computation fault appears. V2 thus moved where \textit{Ghost Fills} fail, from the owner's collateral balance to the callback that delivers tokens to the recipient.

\finding{Removing \texttt{incrementNonce()} in V2 did not reduce \textit{Ghost Fills}; the dominant failure surface shifted to rejected token-delivery callbacks.}

The long tail contains direct evidence that some failures are engineered. A few reverts carry custom strings compiled into wallet code, such as \texttt{GRIEFING\_ACTIVE}, \texttt{poc-v4:\,griefing on}, and \texttt{ghost-fill:\,receiver invalidated}. Each string appears in only a few transactions, yet together they show that some participants revert settlements on purpose. RQ2 separates these deliberate \textit{Ghost Fills}, caused by \textit{Cancellation Attacks}, from the rest.

\subsection{Affected Markets, Users, and Order Patterns}
\label{sec:rq1:affected}

We join each revert to its market through the Gamma API, resolving 99.8\% of reverts. \textit{Ghost Fills} concentrate in fast, recurring markets. The most common tags are \texttt{Crypto}, \texttt{Up or Down}, \texttt{5Min}, each associated with more than one million reverts. The most affected individual markets are five-minute Bitcoin direction markets. High-stakes event markets also appear: \textit{``English Premier League Winner''} alone has 14{,}902 reverts and \$2.8\,M of collateral at risk. It is a NegRisk multi-outcome market, a setting we revisit when analyzing arbitrage-bot hunting in Section~\ref{sec:disc:arbitrage-hunt}.

Reverts concentrate on a minority of accounts without being dominated by a fixed few. The busiest participant appears in 122{,}753 reverts (6.0\%). The top ten participants account for 11.1\% of all reverts, the top 100 for 32.8\%, and the top 1{,}000 for 69.8\%. The remaining reverts spread across a long tail of more than 260{,}000 participants.

\finding{\textit{Ghost Fill} activity is concentrated but broadly distributed. The top 1{,}000 participants account for 69.8\% of reverts, while the remaining 30.2\% spans more than 260{,}000 participants.}

\subsection{Estimated Financial Impact}
\label{sec:rq1:impact}

We measure direct exposure as the collateral leg of the order that failed to settle, converted to USD. Summed over all reverts, \collateralAtRisk failed to settle on-chain. The distribution is heavy-tailed: the median revert risks only \$11 of collateral and the 90th percentile \$505, but the tail runs to single orders above \$1\,M.
The operator cost is directly observable. Every reverted settlement consumes gas, totaling \operatorGasBurned POL over the window, about \$177\,K at the end-of-window POL price\footnote{We value gas at the POL price of \$0.075 on 2026-08-15.~\cite{coinmarketcapPolygonPrevMATIC2026}}. The platform bears this cost even though the settlement fails.
\textit{Ghost Fills} also create a user-visible consistency failure: the interface reports a fill that later disappears. Users have attributed such reversals to platform-side manipulation on social media~\cite{aggyGf,castleGfStill}. The volume measured above shows that users met this failure often.

\begin{answerbox}
\textbf{Answer to RQ1:} \textit{Ghost Fills} are widespread. Over the study window, \tool labels \ghostFillCount reverted settlements involving \ghostFillParticipants participants, with daily reverts peaking at \peakDailyRate of all \texttt{matchOrders} transactions. These failures leave \collateralAtRisk of collateral unsettled and burn \operatorGasBurned POL (about \$177\,K) of operator gas. They concentrate on insufficient collateral balance (52.5\%) and rejected token-delivery callbacks (39.5\%). Across the V1-to-V2 cutover, the dominant surface shifts from the former to the latter.
\end{answerbox}
\section{Answer to RQ2: Cancellation Attack Vectors}
\label{sec:rq2}

\begin{figure}[t]
    \centering
    \includegraphics[width=\textwidth]{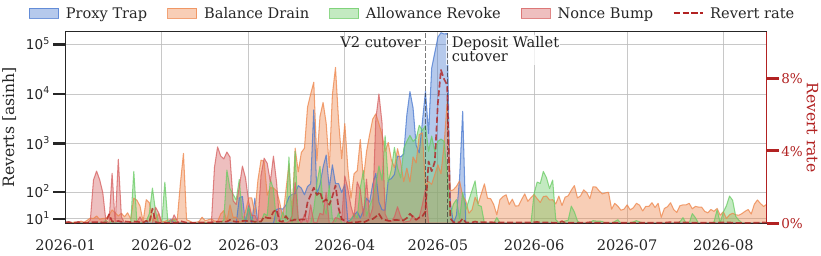}
    \Description{Daily reverted matchOrders from January to August 2026 on an asinh scale, split into Proxy Trap, Balance Drain, Allowance Revoke, and Nonce Bump, with the revert rate overlaid and the V2 cutover and Deposit Wallet cutover marked.}
    \caption{Daily reverted \texttt{matchOrders} attributed to each \textit{Cancellation Attack} vector, with the V1-to-V2 cutover and the Deposit Wallet cutover marked. Nonce Bump disappears at the cutover while the others persist.}
    \label{fig:rq2-reverts-daily}
    \vspace{-1em}

\end{figure}

RQ1 grouped every revert by the on-chain check that failed during settlement. RQ2 asks which of those failures were deliberate cancellations. The four settlement checks in Section~\ref{sec:bg:settlement} read state that a participant still controls after the off-chain match: the owner's nonce, the owner's balance, the allowance granted to the exchange, and the code of the receiving wallet. A participant who changes that state before settlement forces the matched order to revert.
We derive one cancellation vector from each check: Nonce Bump, Balance Drain, Allowance Revoke, and Proxy Trap.
\tool's four rules attribute \attackAttributedCount of the \ghostFillCount \textit{Ghost Fills} (\attackAttributedPct) to \textit{Cancellation Attacks}, placing \attackCollateral of collateral at risk.
These attacks burn \attackGasBurned POL (about \$162\,K) in operator gas, \pct{92} of all operator gas spent on \textit{Ghost Fills}. Figure~\ref{fig:rq2-reverts-daily} shows their daily volume by vector.

\subsection{Evaluation}
\label{sec:rq2:eval}

Following prior works' guidance~\cite{jubaPrincipledSamplingAnomaly2015,liuCharacterizingTransactionrevertingStatements2022}, we validate \tool's classifications by manually inspecting a random sample from each vector, sized at 95\% confidence and a 5\% margin of error: 384 transactions for Proxy Trap (\proxyTrapCount reverts) and Balance Drain (\balanceDrainCount), 380 for Allowance Revoke (\allowanceRevokeCount), and 379 for Nonce Bump (\nonceBumpCount)~\cite{calculator.netSampleSizeCalculator2025}. For each sampled transaction, we examine the on-chain execution trace, the causal transaction in the surrounding blocks, and the wallet bytecode where applicable. Two authors label every sample independently; they disagree on 6 of the 1{,}527 samples, and a third author resolves these. None of the four samples contains a misclassification.

\paratitle{Proxy Trap Validation} We check Proxy Trap rule against wallet state. Reading the fallback-handler slot of all 6{,}927 trapped wallets at their first revert block, we find a custom handler in 93.4\% of them, against 0.35\% (1 of 285; Wilson 95\% CI [0.06\%, 1.96\%]) of participants that were never trapped.
The gas profile separates traps just as cleanly. In gas-exhaustion reverts, the wallet's own \texttt{onERC1155Received} frame consumes a median of 786K gas and never less than 416K, while over 61.0M successful deliveries the same frame never exceeds 21K (p99 10.6K). Because this gas is spent inside the recipient's code, the 19$\times$ gap reflects the wallet itself.

\subsection{Proxy Trap}
\label{sec:rq2:proxy-trap}

Proxy Trap is the dominant vector and surged under V2: \tool attributes \proxyTrapCount of the \attackAttributedCount attributed reverts to it, almost all under V2.
A participant wallet is engineered so that the exchange's ERC1155 receiver callback reverts during settlement, preventing outcome-token delivery and reverting the transaction. In its dominant form the wallet keeps the official Gnosis Safe implementation and replaces only its fallback handler (Section~\ref{sec:method:rules}). The handler \addr{0x36750b9d...} is typical. It burns 1{,}173{,}604 gas and then executes \texttt{revert(0,0)}, so the same trap surfaces as an out-of-gas error or as an empty revert depending on the gas the operator forwards.
We observe it on both roles: 487{,}270 reverts trapped the maker side and 308{,}358 the taker side.
A trapped wallet remains armed across many blocks; the most prolific one produced 1{,}480 reverts, including 589 in a single seven-block window of roughly twelve seconds.

\paratitle{Implementation Variants}
This vector exhibits the greatest implementation diversity, yielding a family of variants whose only shared feature is a reverting callback.
We recover 30 distinct trap implementations from on-chain bytecode and catalog them in Table~\ref{tab:rq2-proxy-variants}, falling into three escalating styles. 
The simplest variants exhaust the gas that the exchange forwards into the callback. The largest (F-1), deployed behind 4{,}148 wallets, has its receiver \texttt{delegatecall} back into itself on every token receipt, recursing without a base case until the entire gas budget is consumed and settlement reverts with \texttt{out of gas}. A related implementation, F-4, instead creates the 1{,}024 nested calls needed to reach the EVM call-stack depth limit. Both leave a clear signature, since honest receiver callbacks never exhaust gas or overflow the call stack.

Later variants hide the trap behind a plausible revert reason. One handler (F-5) reverts with the exchange's own \texttt{OrderFilledOrCancelled()} selector, so an operator inspecting the failure reads it as an ordinary double fill.
Another (S-2) impersonates the OpenZeppelin \texttt{ERC20: transfer amount exceeds balance} string, so the revert looks like an ordinary balance shortfall.

These disguised variants were preceded by a handful of proof-of-concept contracts that name the abuse outright, reverting with strings such as \texttt{ghost-fill:\,receiver invalidated} (F-10) and \texttt{poc-v4:\,griefing on} (F-11).

\begin{figure}[t]
    \centering

\begin{lrbox}{\griefbox}\begin{minipage}{\linewidth}
\begin{lstlisting}[language=Solidity]
contract GriefingHandler {
    bool private armed;

    function setArmed(bool v) external {
        require(msg.sender == address(this));
        armed = v;
    }

    function onERC1155Received(address, address, uint256, uint256, bytes calldata)
        external view returns (bytes4) {
        require(!armed, "GRIEFING_ACTIVE");            // reject delivery if armed on
        return this.onERC1155Received.selector;
    }
}
\end{lstlisting}
\end{minipage}\end{lrbox}
\resizebox{0.85\linewidth}{!}{\usebox{\griefbox}}

\Description{Solidity listing of a GriefingHandler contract whose onERC1155Received callback reverts with GRIEFING_ACTIVE once armed by a self-call.}
\caption{Core of the EIP-7702 griefing implementation: a normal receiver until a self-call arms it.}
\label{fig:griefing}
\vspace{-1em}
\end{figure}

The newest carrier removes the proxy wallet entirely. EIP-7702~\cite{qiEIP7702PhishingAttack2025} lets an externally owned account attach contract code to itself through a type-4 set-code transaction, while keeping the same address, balance, and storage. After delegation, the account has nonzero code size, so the exchange treats it as a contract recipient and invokes the ERC1155 receiver callback.
We observe this pattern in EOA \addr{0x542e...a62a} (D-1). A single type-4 transaction delegates the account to a griefing implementation and arms it. While disarmed, the callback returns the expected ERC1155 magic value, allowing the wallet to trade normally. Once armed through a self-call, the same callback reverts with \texttt{GRIEFING\_ACTIVE}, causing every subsequent token delivery to fail. Figure~\ref{fig:griefing} shows the recovered implementation core decompiled by Dedaub~\cite{grechGigahorseThoroughDeclarative2019a}. This variant extends Proxy Trap to plain EOAs and evades detectors that inspect only contract wallets.

Another variant (F-12) shows that a trap can lie dormant. Thirteen Gnosis Safes were created on May 3 and 4, just before the Deposit Wallet upgrade that defends against this attack, and then sat idle for days. Each installed the F-12 handler one or two blocks before its first revert. Together they caused a spike of 4{,}498 reverts on May 9.

\finding{Proxy Trap is engineered and evolving. We recover 30 implementations, ranging from gas burns and stack overflows to callbacks that mimic benign exchange or ERC-20 errors, and finally to an EIP-7702 variant that traps a plain EOA.}

\paratitle{Market Concentration} Proxy Trap is also the most market-concentrated vector. \pct{99.2} of its reverts land in recurring crypto markets and only \pct{0.1} in NegRisk markets, against a \pct{9.8} NegRisk share across all \textit{Ghost Fills}.
The reverts cluster in the five-minute Bitcoin up-or-down markets, which carry the highest churn on the platform.

\finding{Proxy Trap is also the costliest vector. Its gas-exhaustion variants consume the whole transaction gas budget, so each revert burns 2.7 POL against 0.06 for an ordinary settlement. The vector alone accounts for \proxyTrapGasBurned POL (about \$161\,K), \pct{91.3} of all operator gas spent on \textit{Ghost Fills}, and constitutes a denial-of-service to the settlement system.}

\subsection{Nonce Bump}
\label{sec:rq2:nonce-bump}

Nonce Bump is a V1-only vector: the signer calls \texttt{incrementNonce()} on the exchange, which invalidates the signed nonce of an already-matched order, so the settlement reverts with \texttt{InvalidNonce()}. \tool attributes all \nonceBumpCount \texttt{InvalidNonce} failures from RQ1 to it, across 2{,}701 distinct attacker addresses. V2 removed the function and closed the vector.

The vector escalated into a cat-and-mouse dynamic under V1. The first generation (N-1) broadcast \texttt{incrementNonce()} openly as a top-level EOA transaction, often with a high gas price, so the bump would land just before the operator's settlement. Such top-level calls are trivial to watch for, and Polymarket began monitoring the function and banning the signers who issued them.

The attackers responded by hiding the call inside a wallet transaction. Most hand-craft a Gnosis Safe \texttt{execTransaction} that invokes the exchange directly (N-2); 773 reverts instead route the call through Polymarket's own proxy-wallet factory or a smart-account wallet (N-4). Later attackers buried even that (N-3): the signer issues an \texttt{execTransaction} with \texttt{operation = DELEGATECALL} into the Safe \texttt{MultiSend} helper, whose batched payload carries the bare \texttt{incrementNonce()} selector as one of its inner calls. The bump then runs as an internal call inside an innocuous-looking Safe transaction and is invisible to any monitor that scans only top-level calldata for the selector. We confirmed instances from the proxy wallet \addr{0x9391...8127}, whose bumps appear only inside such nested \texttt{multiSend} batches.

\finding{Nonce Bump runs an arms race: once Polymarket monitors and bans the EOAs that broadcast \texttt{incrementNonce()} openly, attackers hide the same call as an internal call that top-level transaction scanners cannot see.}

Nonce Bump over-indexes on NegRisk markets (\pct{23} of its reverts, against the \pct{9.8} platform baseline), which points to the arbitrage-bot hunting discussed in Section~\ref{sec:disc:arbitrage-hunt}.

\subsection{Collateral Withdrawal}
\label{sec:rq2:collateral}

\definecolor{cVuln}{RGB}{190,50,50}
\definecolor{cGreen}{HTML}{009E73}
\definecolor{cBlue}{HTML}{0072B2}
\definecolor{cYellow}{HTML}{E6A700}

\begin{figure}[t]
\centering
\resizebox{\textwidth}{!}{%
\begin{tikzpicture}[
  obj/.style   ={rounded corners=2.5pt, line width=0.5pt,
                 align=center, inner sep=3pt, text width=2.3cm,
                 minimum height=0.75cm, font=\small\bfseries},
  atk/.style   ={obj, fill=cVuln!10, draw=cVuln!55, text=black},
  fee/.style   ={obj, fill=cYellow!22, draw=cYellow!75!black, text=black},
  lifeline/.style={densely dashed, line width=0.4pt, color=black!40},
  call/.style  ={-{Stealth[length=1.8mm,width=1.4mm]}, line width=0.7pt,
                 color=black!75},
  pending/.style={-{Stealth[length=1.7mm,width=1.3mm]}, line width=0.6pt,
                 densely dashed, color=black!45},
  bad/.style   ={-{Stealth[length=2.0mm,width=1.6mm]}, line width=0.85pt,
                 color=cVuln!75!black},
  badret/.style={-{Stealth[length=1.8mm,width=1.4mm]}, line width=0.8pt,
                 densely dashed, color=cVuln!75!black},
  act/.style   ={font=\scriptsize, text=black!78, align=left,
                 inner sep=1.5pt, anchor=west},
  num/.style   ={font=\fontsize{6.5}{7.6}\selectfont\ttfamily,
                 text=black!55},
  note/.style  ={draw=black!40, line width=0.4pt, fill=black!3,
                 rounded corners=1pt, align=center, inner sep=2.5pt,
                 font=\fontsize{7}{8.4}\selectfont, text=black!70},
  bar/.style   ={fill=black!8, draw=black!45, line width=0.4pt},
]

\def\yG{0}
\def\yP{-1.35}
\def\yF{-2.7}
\def\xL{1.45}
\def\xR{14.3}
\def\h{0.07}

\def\sub#1{{\fontsize{6}{7}\selectfont\ttfamily\mdseries\textcolor{black!55}{#1}}}
\node[atk] (G) at (0,\yG) {Funding Wallet\\[-1pt]\sub{0x4BFE\dots 0b8d}};
\node[atk] (P) at (0,\yP) {Proxy Wallet\\[-1pt]\sub{0x9DD9\dots cAE23}};
\node[fee] (F) at (0,\yF) {Fee Module\\[-1pt]\sub{matchOrders()}};

\foreach \y in {\yG,\yP,\yF}
  \draw[lifeline] (\xL,\y) -- (\xR,\y);

\draw[bar] (1.65,\yP-\h) rectangle (\xR,\yP+\h);
\draw[bar] (4.55,\yF-\h) rectangle (10.7,\yF+\h);
\foreach \xa/\xb in {{1.9}/{2.3}, {7.1}/{7.5}, {12.2}/{12.6}}
  \draw[bar] (\xa,\yG-\h) rectangle (\xb,\yG+\h);

\draw[call] (2.1,\yG-\h) -- (2.1,\yP+\h);
\node[act] at (2.15,{(\yG+\yP)/2})
  {\textbf{1.\ fund order}\\\texttt{272.5\,USDC.e}\\\texttt{gas=604\,Gwei}};

\node[note, anchor=north] at (3.3,\yP-0.18)
  {signs order;\\CLOB matches\\off-chain \textcolor{cGreen}{\faCheckCircle}};

\draw[pending] (4.75,\yF+\h) -- (4.75,\yP-\h);
\node[act, text=black!60] at (4.8,{(\yP+\yF)/2})
  {\textbf{2.\ matchOrders()}\\sent, \emph{queued in mempool}\\\textcolor{black!55}{\texttt{gas 245\,Gwei (low)}}};

\draw[bad] (7.3,\yP+\h) -- (7.3,\yG-\h);
\node[act, text=cVuln!75!black] at (7.35,{(\yG+\yP)/2})
  {\textbf{3.\ drain collateral} \textcolor{cYellow}{\faBolt}\\\texttt{272.5\,USDC.e}\\\texttt{gas 2{,}350\,Gwei (front-run)}};

\draw[call] (9.9,\yF+\h) -- (9.9,\yP-\h);
\node[act, anchor=east, align=right] at (9.85,{(\yP+\yF)/2})
  {\textbf{4.\ matchOrders()} mined\\\texttt{gas=245\,Gwei}\\\textcolor{black!55}{\texttt{try transferFrom()}}};
\draw[cVuln,line width=0.9pt] (9.78,\yP+0.12)--(10.02,\yP-0.12);
\draw[cVuln,line width=0.9pt] (9.78,\yP-0.12)--(10.02,\yP+0.12);

\draw[badret] (10.5,\yP-\h) -- (10.5,\yF+\h);
\node[act, text=cVuln!75!black, font=\fontsize{6.5}{7.6}\selectfont\ttfamily] at (10.55,{(\yP+\yF)/2+0.2})
  {TRANSFER\_FROM\_FAILED};
\node[note, anchor=north west] at (10.65,{(\yP+\yF)/2-0.02})
  {CLOB clears; \textcolor{cVuln}{\textit{Ghost Fill} \faTimesCircle}};

\draw[call] (12.4,\yG-\h) -- (12.4,\yP+\h);
\node[act] at (12.45,{(\yG+\yP)/2})
  {\textbf{5.\ re-fund}\\\textbf{next order}\\\texttt{272.5\,USDC.e}};

\foreach \y in {\yG,\yP,\yF}
  \node[font=\normalsize,text=black!45] at (14.55,\y) {$\cdots$};
\node[font=\fontsize{7}{8.4}\selectfont\itshape,text=cVuln,anchor=north east]
  at (14.8,\yF-0.2) {\faRedo \;\; attack continues \textbf{$\times$369 rounds}};

\end{tikzpicture}%
}
\vspace{-1.5em}
\caption{An instance of collateral withdrawal via Balance Drain.}
\Description{Sequence diagram with time running left to right across three lifelines: funding wallet, proxy wallet, and fee module. The funding wallet funds the proxy, the operator's matchOrders is queued with low gas, the proxy drains its collateral back with a high-gas front-run, the mined matchOrders fails with TRANSFER_FROM_FAILED, and the loop repeats 369 times.}
\label{fig:rq2-balance-drain-seq}
\vspace{-1.5em}
\end{figure}
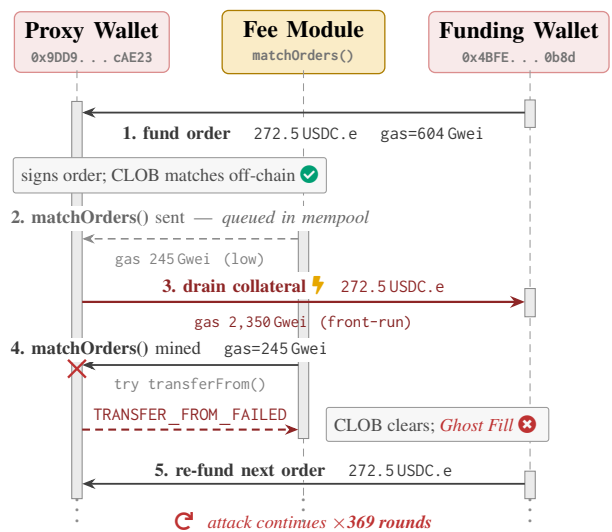

The remaining two vectors both make a participant's collateral unavailable just before settlement, and both survived the V2 cutover. They sit on opposite sides of the trade: Balance Drain is overwhelmingly the taker pulling its own collateral (89\% of its reverts), whereas Allowance Revoke is most often a maker cutting the exchange's allowance (61\%). Each is a front-running race in which the draining or revoking transaction outbids the settlement it cancels. The causal transactions behind \textit{Cancellation Attacks} bid a median 9.7$\times$ the median gas price of their block, against 1.0$\times$ for other transactions in the same blocks, and even their 10th percentile (2.6$\times$) lies above the ordinary 90th percentile (1.9$\times$).

\paratitle{Balance Drain} Balance Drain moves all collateral out of the paying wallet, so the exchange's \texttt{transferFrom} finds an empty balance; \tool attributes \balanceDrainCount reverts to it across 25{,}433 addresses, the widest attacker population of any vector. The proxy wallet \addr{0x9dd9...ae23} is representative of 511 such reverts.
Its signer front-runs each pending settlement with a high-gas \texttt{execTransaction} priced at 2{,}349.6 Gwei against a standard 276\,Gwei~\cite{polygonscanPOLYGasTracker2026}. The transaction lands first in its block and sweeps the wallet's USDC.e~\cite{circleBridgedUSDCStandard2026}, so the lower-gas \texttt{matchOrders} that follows in the same block reverts.
It recycles its collateral through a fixed deposit-drain loop (B-1). Under V1 alone, its funding wallet refills it 369 times~(Figure~\ref{fig:rq2-balance-drain-seq}), and after the cutover a second funding wallet keeps the loop running in pUSD.

A more elaborate variant (B-2) batches the drain across many wallets at once. A single EOA \addr{0x83d4...de5f} empties about twenty funded wallets in one \texttt{Multicall3} \texttt{aggregate3} transaction, placed near the top of its block, that converts their combined 17.7K USDC.e of collateral into DAI through a Uniswap V3 pool, reverting 20 related \texttt{matchOrders} at once.

\paratitle{Allowance Revoke} Allowance Revoke instead shrinks or zeroes the exchange's token allowance, so the same transfer fails one step earlier, on the allowance check; \tool attributes \allowanceRevokeCount reverts to it across 12{,}384 addresses. For instance, the EOA \addr{0x2539...6739} produced 954 reverts this way, 861 of them (\pct{90}) in the same block as the \texttt{matchOrders} they cancelled, almost all by calling \texttt{approve(exchange, 0)} on USDC.e (A-1). Because revoking touches no balance, it needs only a modest gas premium, and the allowance can be reinstated at any time, so the orders keep appearing fillable while none can settle.

The two vectors split along market lines. Balance Drain tracks the binary-crypto baseline, with \pct{89.4} of its reverts in recurring crypto markets and \pct{6.1} in NegRisk. Allowance Revoke over-indexes on NegRisk (\pct{22}, against the \pct{9.8} baseline) and shows the lightest crypto skew of any vector.

\subsection{Benign Reverts}
\label{sec:rq2:benign}

Three recurring failure modes outside the four vectors are platform-side bugs. \textit{1)} The V2 fee check, \texttt{FeeExceedsMaxRate()}, fires on legitimate market-sell orders because the contract validates the fee against the taker's signed limit price instead of the proceeds the match delivers. It reverts a flow the official interface itself produces, and accounts for all 21{,}076 V2 fee-computation reverts in Table~\ref{tab:rq1-failure-surface}.
\textit{2)} NegRisk settlements occasionally run out of gas because each additional maker forces another on-chain split-mint, and gas grows with the maker count.
\textit{3)} A taker can lose a settlement race when its concurrent same-block orders together exceed its balance while each is individually fundable, a TOCTOU gap~\cite{mitreCWE367TimeofcheckTimeofuse2026} in the operator's off-chain balance check; one market-making bot \addr{0x850a...ce8b} alone accumulated 1{,}103 such reverts.
We reported all three to Polymarket through the disclosure in Section~\ref{sec:disc:disclosure}.

\begin{answerbox}
\textbf{Answer to RQ2:} \tool attributes \attackAttributedCount of the \ghostFillCount \textit{Ghost Fills} to \textit{Cancellation Attacks} through four vectors, placing \attackCollateral of collateral at risk. Behind these vectors lie 37 implementation variants evolved against each of Polymarket's patches and monitoring rules.
\end{answerbox}

\section{Answer to RQ3: Spread of Risk}
\label{sec:rq3}

RQ1 and RQ2 measure Polymarket alone, yet the settlement gap is a property of the hybrid exchange design. We therefore measure how widely the design has been reused on other prediction platforms, then examine when reusing this architecture turns the latent gap into an exploitable vulnerability.

\input{figures/rq3_jaccard_barcode}

\subsection{Cross-Chain Reuse of the Exchange}
\label{sec:rq3:reuse}

Reuse of Polymarket's design is common and long-tailed. Across all verified contracts on 401 chains, 31{,}897 share at least one function selector with a Polymarket contract. Of these, \pct{96} score a Jaccard below 0.2, the level that common ERC-20 and ERC-1155 templates produce on their own, and 494 reach 0.5 or higher. We cut at 0.5 and manually inspect the remaining contracts to exclude coincidental matches, testnet deployments, and a handful of duplicate or Polymarket-owned addresses. Dropping these leaves \reusedContracts independent reuses across \reuseChainCount chains, concentrated on Base and BSC. Similar selectors do not by themselves imply similar behavior, so we also read the verified source of all \reusedContracts contracts. Every one keeps Polymarket's settlement logic, which re-checks each order against live on-chain state only when \texttt{matchOrders} executes, and thus inherits the same gap. \jaccardOneCount of them still score a perfect 1.0, reproducing an official exchange's interface down to its original contract names, \texttt{CTFExchange}, \texttt{NegRiskFeeModule}.

Figure~\ref{fig:rq3-jaccard-barcode} shows a similarity comparison example. Many are in production: \reuseActive remained active past the end of our window and average \reuseAvgTx settled matches each, the largest a fee module on Base carrying 5.1M.
Attributing the deployments to operating entities shows that several prediction-market projects reuse Polymarket's exchange design (Table~\ref{tab:rq3-forks}). The \textit{Ghost Fill} rates in the table are peak daily revert rates, computed as reverted \texttt{matchOrders} transactions divided by all \texttt{matchOrders} transactions on that day.

Three active prediction-market reuses still show \textit{Ghost Fills}: P-1, P-2, and P-4. P-1 has moved from plain EOAs to an ERC-4337 EntryPoint and an officially designated smart-account model. This setup mitigates the main cancellation vectors, and its observed daily revert rate is only \pct{0.2}. P-4 reproduces Polymarket's structure more closely and ghosts roughly \pct{1.0} of daily matches; its NegRisk path also exhibits the same out-of-gas failure discussed in Section~\ref{sec:rq2:benign}. P-2 is the most affected active fork in our sample: on 2026-05-12, its CTF Exchange processed 6{,}448 \texttt{matchOrders} transactions, of which 1{,}104 reverted (\pct{17.1}). Its own audit also identifies the \texttt{incrementNonce} denial-of-service vector we classify in Section~\ref{sec:rq2:nonce-bump}.
P-5, an earlier form of P-1 that is no longer active, also shows \textit{Ghost Fills} and marks the design lineage. P-3 documents the possibility of \textit{Ghost Fills} in its own materials.

\finding{The flawed hybrid exchange has become a reusable template. It is live on \reusedContracts independent deployments across \reuseChainCount chains holding at least \$23\,M in user funds, \jaccardOneCount of them exposing an identical function-selector set, and we confirm \textit{Ghost Fills} on four of them, including the two largest.}

\begin{table}[t]
  \centering
  \caption{Reuses of the Polymarket exchange design.}
  \vspace{-1em}
  \label{tab:rq3-forks}
  \footnotesize
  \setlength{\tabcolsep}{3.5pt}
  \begin{tabular}{l l l r l}
    \toprule
    Platform & Category & Chain & TVL & Ghost Fills \\
    \midrule
    P-1     & Prediction & BSC  & \$1*.**\,M & Yes ($\sim$\pct{0.2}) \\
    P-2     & Prediction & BSC & \$7.**\,M & Yes ($\sim$\pct{17.1})$^{\ddagger}$ \\
    P-3     & Prediction & SX   & \$0.6*\,M & Potential$^{\dagger}$ \\
    P-4     & Prediction & Base & \$0.6*\,M & Yes ($\sim$\pct{1.0}) \\
    P-5     & Prediction & BSC  & \$0.2*\,M & Yes ($\sim$\pct{8.9}) \\
    NFT-1   & NFT market & Eth. & \$1.**\,M & None \\
    \bottomrule
  \end{tabular}

  \vspace{2pt}
  {\footnotesize $^{\ddagger}$ Its own audit report identifies the Nonce Bump griefing vector. $^{\dagger}$ Its documentation admits potential \textit{Ghost Fill} patterns.}
  \vspace{-1.5em}

\end{table}

\subsection{From Reuse to Exploitation}
\label{sec:rq3:incentive}

Reuse alone does not explain exploitation. Polymarket's exchange follows the Wyvern-style hybrid model~\cite{lesavreBlockchainNetworksToken2024}: users sign orders off-chain, and an on-chain \texttt{matchOrders} call settles the matched orders later. This design has been used since 2018 and still appears in NFT marketplaces such as NFT-1. In that setting, the same settlement gap exists, but it rarely creates a profitable cancellation strategy.
Prediction markets change the incentives around the same architecture. NFT markets tolerated reversible fills because prices rarely move between matching and settlement. In a prediction market, outcome information can turn a matched order's expected value negative before it settles, so the same gap becomes a profitable cancellation primitive.

\finding{A proven architecture can become unsafe when reused under different economic semantics, turning a long-tolerated consistency gap into an exploitable cancellation vector.}

The forks inherit the same cancellation vectors without Polymarket's attack peaks. They lack the liquidity, trading frequency, and per-match rewards (Section~\ref{sec:disc:liquidity-rewards}) that make cancellation profitable at scale. Should they grow into the same incentive regime, the same vectors can activate.

\begin{answerbox}
\textbf{Answer to RQ3:} The flawed exchange design is reused by \reusedContracts independent contracts across \reuseChainCount chains holding at least \$23\,M in user funds, \jaccardOneCount of them selector-identical copies, with \textit{Ghost Fills} on four competing prediction markets. The risk extends beyond Polymarket, and any such deployment becomes exposed once the payoff from cancelling grows large enough.
\end{answerbox}

\section{Discussion}
\label{sec:discussion}

\subsection{Profit Measurement}
\label{sec:disc:weaponization}

Forcing a settlement to revert pays off in several ways. We infer three strategies from the markets that attackers target and from reports on social media~\cite{saidRewardFarm,itslirratoSomeoneDrainingNegrisk2026}. 

\paratitle{P1. Risk-Free Prediction}
This strategy runs the motivating example (Section~\ref{sec:bg:motivating-example}) at scale in the five-minute crypto up/down markets. An actor places an order just before resolution and waits. If it lands in the money, the match settles and pays out; otherwise, the actor cancels and the losing fill ghosts, so no position is ever held at a loss. The clearest case we find is \addr{0xc393...b18f}. All 360 markets it trades are five-minute crypto markets, and \tool records 120 \textit{Cancellation Attacks} in them. Of the 287 positions it lets settle, 286 win (99.7\%). Every match it involves is timestamped after the outcome has already been revealed.

\paratitle{P2. Arbitrage-Bot Hunting}
\label{sec:disc:arbitrage-hunt}
A second strategy turns the \textit{Cancellation Attack} against arbitrage bots that patrol Polymarket for mispriced complementary outcomes. In a women's short-track speed-skating market~\cite{polymarketWinterGames20262026}, an actor ran a cluster of ten Sybil wallets, each posting a fake arbitrage opportunity across the mutually exclusive sub-markets to lure a bot into building a position. The actor then cancelled, forcing the bot to hedge and dump its remaining inventory at a loss into a low bid that \addr{0x6e7e...7282} had rested in advance; that wallet redeemed the position after settlement for \$4{,}415.37.

\paratitle{P3. Liquidity-Reward Manipulation}
\label{sec:disc:liquidity-rewards}
Polymarket offers an official liquidity-rewards program for liquidity providers~\cite{polymarketLiquidityRewards2026}. This incentive enables a third manipulation strategy. The maker posts a reward-eligible quote, accrues a reward score, and cancels as soon as the quote is matched. It therefore earns the payout without taking real inventory risk.
We confirm it by cross-referencing attacker addresses with the on-chain reward distributor: \$13{,}397.65 in liquidity rewards flowed to 1{,}567 attacker addresses. The clearest case is \addr{0x0fb5...d1a0}. This address triggered 466 Proxy Trap reverts over three days, settled no successful fills, and still received \$4{,}425.88 from the reward program. Because it realized no trading profit and completed no match, the payout appears to be entirely farmed reward.

\paratitle{Avoided Loss}
P1 to P3 share one goal: the cancelling party avoids a loss that a settled fill would lock in. We therefore measure the loss that each \textit{Cancellation Attack} avoids. For each attack-labeled revert, we take the order parameters in its calldata as the quantity that would have filled, and value that fill with the market resolution reported by the Gamma API~\cite{polymarketGammaAPI2026}. A buy that pays $c$ USD for $q$ shares of an outcome with payout $x$ avoids a loss of $c - qx$; a sell avoids $qx - c$.

The \attackAttributedCount attack reverts collapse into \uniqueOrders unique signed orders, since the same order can be matched and reverted many times. Their notional value is \$141.3\,M. If these orders were settled, their owners would have lost \avoidedLoss, or 73 cents per dollar of notional.

\subsection{Organized Exploitation}
\label{sec:disc:organized}

The attacker accounts are not the independent users they appear to be on-chain. \tool attributes \textit{Cancellation Attacks} to 47{,}441 distinct addresses, yet 65.3\% of them appear in a single revert before going dormant. Address-association analysis traces many of these wallets to shared funders and deployers that mass-produce them: the funder \addr{0xbf1d...a492} seeded 24{,}539 fresh wallets in a five-day burst at nearly one per minute, and the account \addr{0xd4bc...9f13} deployed 792 Gnosis Safe proxy wallets across 3{,}333 transactions in three days.

The funding behind these wallets shows the same hand. Most funds trace to centralized-exchange and DEX wallets, where the trail ends at the custodial boundary; a minority is laundered deliberately. We observe peel chains~\cite{kapposHowPeelMillion2022}, as shown in Figure~\ref{fig:peel-chain}. A hub receives a round sum, forwards to each throwaway wallet only the gas needed for one cancellation, recovers the unspent change after the attack settles, and passes the remainder to the next hub. No wallet is left holding a balance worth clustering. The same actors register vanity addresses that share leading and trailing bytes and seed one-wei dust between the look-alikes, crowding the address graph with near-duplicates. Others route the gas through the Tornado Cash mixer and successive hops to sever the trail before it reaches an attacker.

\finding{Attackers sustain the campaign against Polymarket's monitoring on two fronts: the attack code keeps evolving past each new detection rule (Section~\ref{sec:rq2}), and tens of thousands of single-use Sybil addresses~\cite{douceurSybilAttack2002} make per-account blacklisting ineffective.}

\subsection{Responsible Disclosure}
\label{sec:disc:disclosure}
\label{sec:disc:mitigation}

We disclosed our findings to Polymarket in three rounds before, during, and after the V2 cutover. We also contacted the third-party deployers from Section~\ref{sec:rq3} whose contracts reuse the vulnerable design, warning each of the exposure it inherits.

Polymarket has answered \textit{Ghost Fills} with two upgrades. The V2 cutover~\cite{polymarketMigratingCLOBV22026} removed \texttt{increment\allowbreak Nonce()}, eliminating Nonce Bump but leaving the other vectors intact, and \textit{Cancellation Attacks} spiked in the week that followed. Then Polymarket shipped the Deposit Wallet~\cite{polymarketDepositWallets2026}, which routes a participant's collateral through a platform-controlled wallet and revokes the participant's authority to move funds or change allowances before settlement. The rollout came with a campaign to blacklist the attacker accounts still active.
Both upgrades leave the structural cause untouched, so \textit{Ghost Fills} keep surfacing. We detail the mitigations we proposed to Polymarket in Appendix~\ref{app:mitigation}.

\section{Threats to Validity}
\label{sec:disc:validity}

\paratitle{Internal Validity}
\label{sec:disc:validity:internal}
We value collateral and gas in USD at \studyEnd{} prices, which may differ from their value at the time of each event. We do not report per-address attacker profit, because each strategy hides its gain where on-chain data cannot fully reach. The gain often lands on companion addresses that we cannot link to the cancelling one, and it mixes with ordinary trading or honest quoting. Tracing it fully would also require the CLOB order flow and the reward scores that Polymarket computes off-chain.

\paratitle{External Validity}
\label{sec:disc:validity:external}
Our measurement covers one venue over one year. \textit{Ghost Fills} are rare before 2026 (Figure~\ref{fig:rq1-reverts-daily}), so the period before the Fee Module went live adds little. Our measurement rests on public on-chain events, so cancellations that take effect inside Polymarket's private CLOB before any settlement is submitted fall outside its scope. The cross-chain analysis in Section~\ref{sec:rq3} covers only Sourcify-verified contracts, so closed-source reuse remains unmeasured.

\section{Related Work}
\label{sec:related}

\subsection{Prediction Market Measurement}
\label{sec:related:empirical}
Saguillo \etal~\cite{saguilloUnravellingProbabilisticForest2025} study arbitrage on Polymarket and separate market-rebalancing arbitrage from combinatorial arbitrage, estimating roughly \$40M in extracted profit.
Tsang \etal~\cite{tsangAnatomyBlockchainPrediction2026} reconstruct an election market from on-chain data and distinguish genuine trading volume from share creation and destruction; they report that arbitrage inefficiencies shrank from hours to under a minute as the market matured.
Jia \etal~\cite{jiaUnlockingForecastingEconomy2026} release a lifecycle dataset that spans more than 770K market records, 943M trades, and 2M oracle events.
These works focus on how markets price, trade events, and resolve via on-chain reported fill. We study a different, adversarial layer: whether an order matched off-chain actually settles on-chain, and how attackers can force that settlement to fail.

\subsection{MEV and Front-Running}
\label{sec:related:mev}
Daian \etal~\cite{daianFlashBoys202020} introduce maximal extractable value (MEV) and show that arbitrage bots front-run ordinary trades on decentralized exchanges through priority gas auctions, turning transaction ordering into a consensus-layer security risk.
Torres \etal~\cite{torresFrontrunnerJonesRaiders2021} measure front-running on Ethereum at scale and separate it into displacement, insertion, and suppression.
On the defensive side, Yang \etal~\cite{yangSoKMEVCountermeasures2024} systematize thirty MEV countermeasures and study the deployed auction-based solutions empirically, finding that they introduce censorship of their own.
These attacks extract value by reordering or inserting transactions in a single on-chain execution environment.
\textit{Ghost Fills} exploit a different boundary: the attacker does not reorder honest transactions but cancels its own already-matched order, destroying value agreed off-chain by forcing its on-chain settlement to revert.
Ordering helps against some vectors, yet Proxy Trap tampers with the wallet's own implementation so that settlement fails regardless of where the transaction lands.

\subsection{Smart Contract Security}
\label{sec:related:appsec}
Nyx~\cite{zhangNyxDetectingExploitable2024} formalizes exploitable contract-layer front-running vulnerabilities and detects them statically, pruning unrelated function pairs on a hybrid flow graph before validating with symbolic execution that a malicious user can front-run a victim for profit.
SAILFISH~\cite{boseSAILFISHVettingSmart} targets state-inconsistency bugs such as transaction-order dependence, pairing a lightweight exploration phase with symbolic refinement to flag forty-seven vulnerable contracts on Etherscan.
Liu \etal~\cite{liuCharacterizingTransactionrevertingStatements2022} characterize the transaction-reverting statements that Ethereum contracts use to abort anomalous transactions, measuring the security impact of removing each guard.
These studies stay within a single layer: a flaw in the contract code.
The \textit{Cancellation Attacks} we study arise from a gap at the off-chain/on-chain boundary between off-chain fills and on-chain settlement. Analyzing contracts or transactions in isolation cannot reveal them.
\section{Conclusion}
\label{sec:conclusion}

With \tool we labeled \ghostFillCount{} \textit{Ghost Fills} on Polymarket and showed that beneath them runs a deliberate \textit{Cancellation Attack}: an actor voids an order it has already matched by forcing its on-chain settlement to revert. \tool attributes \attackAttributedCount{} of the reverts (\attackAttributedPct) to such attacks, across four vectors and 37 variants that evolved against each defense Polymarket deployed, sparing their owners \avoidedLoss in net losses, burning about \$162\,K of operator gas, and at peak reverting \peakHourlyRate of \texttt{matchOrders} transactions in a single hour. The same flawed exchange is reused by \reusedContracts{} deployments across \reuseChainCount{} chains holding at least \$23\,M in user funds. Because that gap is structural, Polymarket's upgrades mitigate it without curing it.

\bibliographystyle{ACM-Reference-Format}
\bibliography{reference}

\appendix

\section{Implementation Details of \tool}
\label{app:tool-details}

\begin{table}[h]
  \centering
  \caption{Polymarket contracts measured in this study.}
  \label{tab:contracts}
  \footnotesize
  \setlength{\tabcolsep}{4pt}
  \begin{tabular}{l l l l}
    \toprule
    Gen. & Contract & Address & Official Period \\
    \midrule
    V1 & Fee Module                   & \texttt{0xe3f18a...} & 08/15/25 -- 04/28/26 \\
    V1 & NegRisk Fee Module$^{\dagger}$ & \texttt{0xb76889...} & 08/15/25 -- 04/28/26 \\
    V1 & CTF Exchange                 & \texttt{0x4bfb41...} & 08/15/25 -- 04/28/26 \\
    V1 & NegRisk CTF Exchange         & \texttt{0xc5d563...} & 08/15/25 -- 04/28/26 \\
    V2 & CTF Exchange                 & \texttt{0xe11118...} & 04/28/26 -- \\
    V2 & NegRisk CTF Exchange         & \texttt{0xe2222d...} & 04/28/26 -- \\
    \bottomrule
  \end{tabular}

  \vspace{2pt}
  {\footnotesize $^{\dagger}$ The NegRisk Fee Module was upgraded once during the V1 era, which is labeled ``Fee Module V2'' by Polymarket, while it is still a V1-generation contract.}
\end{table}

\tool is implemented in Python and processes one reverted \texttt{matchOrders} transaction at a time. It replays each transaction through a batched \texttt{eth\_call} against an archive node, decodes the resulting custom-error, \texttt{Error(string)}, or \texttt{Panic(uint256)} payload against a selector-to-name map built from the official contract sources, and reconstructs the internal call trace and state diff. It accepts a replayed trace only when its gas use matches the gas the chain recorded, because a node with missing archive state can replay a different execution. The per-vector rules run as priority-ordered plug-ins (Algorithm~\ref{alg:classify}) in the order \texttt{proxy\_trap} $>$ \texttt{nonce\_bump} $>$ \texttt{allowance\_revoke} $>$ \texttt{balance\_drain}; the highest-priority rule whose guard and causal facts both hold claims the rare revert that more than one rule could explain. Table~\ref{tab:contracts} lists the settlement contracts \tool scans: the four \texttt{matchOrders} endpoints across V1 and V2, together with the two underlying V1 exchanges that hold the per-owner nonce.

\begin{algorithm}
\begin{algorithmic}[1]
\Procedure{Classify}{$t$; $R$}
    \State \textbf{input:} reverted \texttt{matchOrders} $t$; rules $R$ by priority
    \State \textbf{output:} a labeled \textit{Ghost Fill} record, or \textsc{unclassified}
    \State $\Phi \gets \textsc{Skeleton}(t) \cup \textsc{Frames}(t)$
        \Comment{skeleton + frame facts}
    \For{rule $\rho \in R$ in ascending priority}
        \If{$\rho.\textsc{guard}(\Phi)$}
            \State $\Phi \gets \Phi \cup \textsc{Probe}(\rho, t)$
                \Comment{causal facts}
            \If{$\rho.\textsc{body}(\Phi)$}
                \State \textbf{return} $\rho.\textsc{emit}(\Phi)$
            \EndIf
        \EndIf
    \EndFor
    \State \textbf{return} \textsc{unclassified}
\EndProcedure
\end{algorithmic}
\caption{\tool's \textit{Ghost Fill} classify algorithm.}
\label{alg:classify}
\end{algorithm}

\section{Cancellation Attack Measurements}
\label{app:rq2-tables}

Table~\ref{tab:rq2-vectors} reports, for each vector, the reverted settlements split across V1 and V2, the count of distinct attacker addresses, and the collateral at risk in millions of USD. Table~\ref{tab:rq2-proxy-variants} catalogs the 37 implementation variants behind the four vectors. Proxy Trap variants are distinct on-chain bytecodes grouped by carrier: a malicious fallback handler (F), a swapped proxy singleton (S), and a direct contract or EIP-7702-delegated EOA (D). The other three vectors are distinguished instead by how the cancelling action is delivered. Event counts are reverted settlements over V1 and V2 combined, each group row carries the per-vector total, and Proxy Trap's eleven low-volume handlers are pooled in one row.

\begin{table}[t]
  \centering
  \caption{\textit{Cancellation Attack} vectors attributed by \tool: reverted
  settlements (V1 / V2), distinct attacker addresses, and collateral at risk (M).}
  \label{tab:rq2-vectors}
  \footnotesize
  \setlength{\tabcolsep}{3.2pt}
  \begin{tabular}{l r r r r r}
    \toprule
    Vector & V1 & V2 & Total & Attkr. & \$\,M \\
    \midrule
    Proxy Trap       & 47{,}524  & 748{,}104 & 795{,}628 & 6{,}927  & 847.2 \\
    Balance Drain    & 127{,}844 & 15{,}218  & 143{,}062 & 25{,}433 & 496.0 \\
    Allowance Revoke & 20{,}792  & 8{,}924   & 29{,}716  & 12{,}384 & 79.6 \\
    Nonce Bump       & 25{,}790  & 0         & 25{,}790  & 2{,}701  & 36.3 \\
    \midrule
    Total            & 221{,}950 & 772{,}246 & 994{,}196 & ---      & 1{,}459.1 \\
    \bottomrule
  \end{tabular}
\end{table}
\begin{table}[t]
  \centering
  \caption{Cancellation-attack implementation variants, grouped by vector. Group rows give the reverts attributed to each vector.}
  \label{tab:rq2-proxy-variants}
  \footnotesize
  \setlength{\tabcolsep}{4pt}
  \begin{tabular}{l l l r}
    \toprule
    ID & Carrier & Signature / mechanism & Events \\
    \midrule
    \multicolumn{3}{l}{\textit{Proxy Trap} (30 variants)} & \proxyTrapCount \\
    \addlinespace[1pt]
    F-1  & handler   & \texttt{out of gas} (recursive \texttt{delegatecall}) & 365{,}660 \\
    F-2  & handler   & reentrancy-sentry \texttt{out of gas}      & 222{,}002 \\
    F-3  & handler   & \texttt{CustomError(0x92bbf6e8)}           & 88{,}787 \\
    S-1  & singleton & bare \texttt{Reverted}                     & 78{,}649 \\
    F-4  & handler   & \texttt{stack limit reached 1024}          & 28{,}071 \\
    F-5  & handler   & \texttt{OrderFilledOrCancelled()}          & 5{,}214 \\
    F-12 & handler   & gas burn, then bare \texttt{Reverted}     & 4{,}716 \\
    F-6  & handler   & \texttt{Error('invalidated')}              & 1{,}353 \\
    F-7  & handler   & \texttt{Error('rejected')} / \texttt{'Rejected'} & 278 \\
    F-8  & handler   & \texttt{CustomError(0x1663f706)}           & 217 \\
    D-2  & EIP-7702  & 7 forged exchange errors (shared code)     & 215 \\
    F-9  & handler   & \texttt{Error('HARD\_REJECT')}             & 10 \\
    S-2  & singleton & \texttt{ERC20: transfer amount exceeds \dots} & 6 \\
    D-1  & EIP-7702  & \texttt{GRIEFING\_ACTIVE} / not whitelisted & 7 \\
    F-10 & handler   & \texttt{ghost-fill:\,receiver invalidated} & 4 \\
    F-11 & handler   & \texttt{poc-v4:\,griefing on}              & 1 \\
    \dots & \dots    & other 11 low-volume handlers               & 438 \\
    \midrule
    \multicolumn{3}{l}{\textit{Nonce Bump} (4 variants)} & \nonceBumpCount \\
    \addlinespace[1pt]
    N-1  & EOA        & top-level \texttt{incrementNonce()} broadcast & \\
    N-2  & Safe       & \texttt{execTransaction} to exchange          & \\
    N-3  & Safe       & nested \texttt{multiSend} \texttt{delegatecall} & \\
    N-4  & proxy / SA & factory \texttt{proxy()} or smart-account call & \\
    \midrule
    \multicolumn{3}{l}{\textit{Balance Drain} (2 variants)} & \balanceDrainCount \\
    \addlinespace[1pt]
    B-1  & EOA        & per-wallet deposit-drain loop                 & \\
    B-2  & Multicall3 & batched cross-wallet sweep                    & \\
    \midrule
    \multicolumn{3}{l}{\textit{Allowance Revoke} (1 variant)} & \allowanceRevokeCount \\
    \addlinespace[1pt]
    A-1  & EOA        & \texttt{approve(exchange, 0)} pre-settlement   & \\
    \bottomrule
  \end{tabular}
\end{table}

\section{Organized Exploitation}
\label{sec:app:peel-chain}

\begin{figure}
    \centering
    \includegraphics[width=\textwidth]{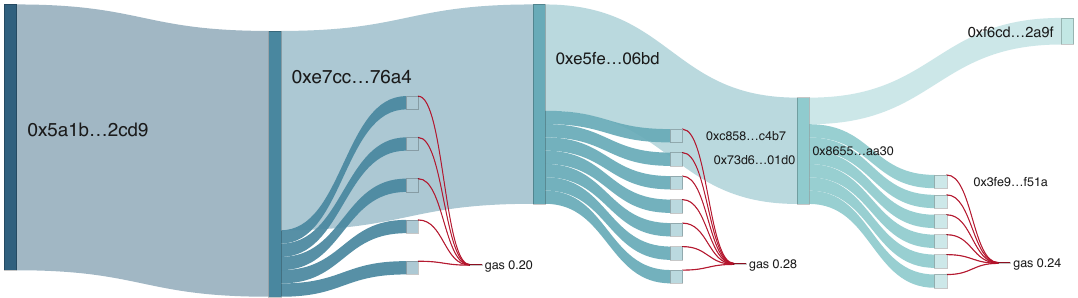}
    \Description{Flow diagram of a funding hub that forwards gas to many throwaway attacker wallets and collects the unspent change back.}
    \caption{Peel-chain laundering of attacker gas: a funding hub forwards each throwaway wallet exactly one cancellation's worth of gas and reclaims the unspent change after the attack settles.}
    \label{fig:peel-chain}
\end{figure}

Figure~\ref{fig:peel-chain} traces the gas funding behind one attacker cluster, showing how a hub seeds throwaway wallets and reclaims the unspent change so that no wallet ever holds a balance worth clustering.

\section{Mitigation Recommendations}
\label{app:mitigation}

We proposed the following mitigations to Polymarket during the disclosure (Section~\ref{sec:disc:disclosure}).

\paratitle{Platform-side recommendations}
The tension between fast matching and on-chain settlement is one of ordering. Any design that reports a fill before its settlement preconditions are irrevocably committed inherits this gap, and committing them at match time closes it. A durable fix must therefore shrink the window between match and settlement that every vector exploits. Short of that, Polymarket can close the TOCTOU gap by escrowing an order owner's collateral at order placement and reserving it against the open order, so the book matches only within already-locked capacity. The more complete remedy attacks the window itself. Polygon is a general-purpose chain whose block time fixes that window and exposes settlement to congestion the platform does not control. Moving settlement to a dedicated, performance-tuned chain would shrink the window until matching and settlement are effectively simultaneous.

\paratitle{Participant-side recommendations}
The arbitrage bots, market-making bots, and AI agents that act on a reported match before it settles are the parties most exposed to \textit{Ghost Fills}. Until that gap closes, they can defend themselves: treat a CLOB fill as provisional and confirm it on-chain before building on it, subscribing to the mempool and waiting for the settling \texttt{matchOrders} to be mined, just as an exchange waits for block confirmations on a deposit. Keeping a shared blacklist of accounts with a history of cancellations further lets them refuse the counterparties most likely to ghost.

\section{BigQuery Queries}
\label{app:sql}

We collect the primary revert dataset with the query in Listing~\ref{lst:sql-reverts}, shown for the V1 contracts; the V2 query differs only in the contract address set, the time window (2026-04-28 to 2026-08-15), and the \texttt{matchOrders} selector (\texttt{0x3c2b4399}). Listing~\ref{lst:sql-jaccard} computes the per-reference selector-set Jaccard score that drives the cross-chain reuse analysis of Section~\ref{sec:rq3}.

\begin{lstlisting}[language=SQL,basicstyle=\ttfamily\scriptsize,captionpos=b,caption={Reverted \texttt{matchOrders} collection (V1).},label={lst:sql-reverts}]
SELECT
  b.block_number,
  r.to_address AS contract_address,
  t.transaction_hash,
  t.block_timestamp,
  t.transaction_index,
  t.input AS tx_input,
  r.gas_used,
  r.effective_gas_price,
  CAST(r.gas_used AS BIGNUMERIC)
    * CAST(r.effective_gas_price AS BIGNUMERIC) AS gas_fee_wei
FROM `bigquery-public-data.goog_blockchain_polygon_mainnet_us.receipts`     AS r
JOIN `bigquery-public-data.goog_blockchain_polygon_mainnet_us.transactions` AS t
  ON r.transaction_hash = t.transaction_hash
JOIN `bigquery-public-data.goog_blockchain_polygon_mainnet_us.blocks`       AS b
  ON r.block_hash = b.block_hash
WHERE r.block_timestamp >= TIMESTAMP('2025-08-15 00:00:00+00')
  AND r.block_timestamp <  TIMESTAMP('2026-04-28 00:00:00+00')
  AND r.to_address IN (
        '0xb768891e3130f6df18214ac804d4db76c2c37730',
        '0xe3f18acc55091e2c48d883fc8c8413319d4ab7b0'
      )
  AND r.status = 0
  AND STARTS_WITH(t.input, '0x2287e350')
ORDER BY r.to_address, b.block_number, t.transaction_index;
\end{lstlisting}

\begin{lstlisting}[language=SQL,basicstyle=\ttfamily\scriptsize,captionpos=b,caption={Cross-chain selector-set Jaccard similarity against the four reference Polymarket exchanges.},label={lst:sql-jaccard}]
WITH
ref_compilations AS (
  SELECT DISTINCT vc.compilation_id
  FROM `sourcify_dataset.public_contract_deployments` cd
  JOIN `sourcify_dataset.public_verified_contracts` vc
    ON vc.deployment_id = cd.id
  WHERE cd.chain_id = 137
    AND cd.address IN (
      FROM_HEX('4bfb41d5b3570defd03c39a9a4d8de6bd8b8982e'),
      FROM_HEX('c5d563a36ae78145c45a50134d48a1215220f80a'),
      FROM_HEX('e3f18acc55091e2c48d883fc8c8413319d4ab7b0'),
      FROM_HEX('b768891e3130f6df18214ac804d4db76c2c37730')
    )
),
ref_sigs AS (
  SELECT DISTINCT rc.compilation_id AS ref_id, ccs.signature_hash_32
  FROM ref_compilations rc
  JOIN `sourcify_dataset.public_compiled_contracts_signatures` ccs
    ON ccs.compilation_id = rc.compilation_id
  WHERE ccs.signature_type = 'function'
),
ref_card AS (SELECT ref_id, COUNT(*) AS n_ref FROM ref_sigs GROUP BY ref_id),
inter AS (
  SELECT ccs.compilation_id AS cid, rs.ref_id,
         COUNT(DISTINCT ccs.signature_hash_32) AS n_inter
  FROM `sourcify_dataset.public_compiled_contracts_signatures` ccs
  JOIN ref_sigs rs USING (signature_hash_32)
  WHERE ccs.signature_type = 'function'
  GROUP BY cid, rs.ref_id
),
cand_card AS (
  SELECT ccs.compilation_id AS cid, COUNT(DISTINCT ccs.signature_hash_32) AS n_cand
  FROM `sourcify_dataset.public_compiled_contracts_signatures` ccs
  WHERE ccs.signature_type = 'function'
    AND ccs.compilation_id IN (SELECT DISTINCT cid FROM inter)
  GROUP BY cid
),
jaccard AS (
  SELECT i.cid,
         MAX(SAFE_DIVIDE(i.n_inter, cc.n_cand + rcd.n_ref - i.n_inter)) AS max_jaccard,
         MAX(i.n_inter) AS best_overlap_count
  FROM inter i
  JOIN cand_card cc ON cc.cid = i.cid
  JOIN ref_card  rcd ON rcd.ref_id = i.ref_id
  GROUP BY i.cid
)
SELECT
  cd.chain_id,
  CONCAT('0x', LOWER(TO_HEX(cd.address))) AS address,
  cc.name AS contract_name,
  cc.fully_qualified_name,
  ROUND(j.max_jaccard, 4) AS max_jaccard,
  j.best_overlap_count,
  vc.created_at AS verified_at
FROM jaccard j
JOIN `sourcify_dataset.public_verified_contracts` vc
  ON vc.compilation_id = j.cid
JOIN `sourcify_dataset.public_contract_deployments` cd
  ON cd.id = vc.deployment_id
JOIN `sourcify_dataset.public_compiled_contracts` cc
  ON cc.id = j.cid
ORDER BY max_jaccard DESC;
\end{lstlisting}

\section{Ethical Considerations}
\label{app:ethics}
This study analyzes only publicly available on-chain data (transaction receipts, contract bytecode, and event logs) and publicly accessible off-chain API data from Polymarket's CLOB. No private user data, off-chain order-book snapshots, or proprietary systems were accessed. All blockchain addresses used in the analysis are pseudonymous identifiers recorded on a public ledger; we do not attempt to link them to real-world identities.

We did not create, deploy, or execute any attack. The four cancellation vectors we document were already being exploited in the wild before our study began; our contribution is to measure and classify existing activity, not to introduce new techniques. The detection rules in \tool operate entirely on historical data and do not interact with live contracts or pending transactions.

We disclosed our findings to Polymarket and tried our best to contact the third-party deployers identified in Section~\ref{sec:rq3} whose contracts reuse the vulnerable exchange design; and the entity identities are also redacted.

\end{document}